\documentclass[twocolumn,tighten]{aastex701}

\usepackage{subcaption}
\usepackage{placeins} 
\usepackage{graphicx}
\usepackage{amsmath}  

\begin{document}

\title{Analysis of synthetic O\,\textsc{VI} absorption associated with galaxy groups in SIMBA and TNG50 simulations}

\author[orcid=0009-0001-5959-9105, gname=Tanmay, sname=Singh]{Tanmay Singh}
\affiliation{School of Earth and Space Exploration, Arizona State University, 781 Terrace Mall, Tempe, AZ 85287, USA}
\email[show]{tsingh65@asu.edu}
\correspondingauthor{Tanmay Singh}

\author[orcid=0000-0002-2724-8298, gname=Sanchayeeta, sname=Borthakur]{Sanchayeeta Borthakur}
\affiliation{School of Earth and Space Exploration, Arizona State University, 781 Terrace Mall, Tempe, AZ 85287, USA}
\email{sborthakur@asu.edu}

\author[orcid=0000-0001-8421-5890, gname=Dylan, sname=Nelson]{Dylan Nelson}
\affiliation{Universität Heidelberg, Zentrum für Astronomie, ITA, Albert-Ueberle-Str. 2, 69120 Heidelberg, Germany}
\email{dnelson@uni-heidelberg.de}

\author[orcid=0000-0003-2842-9434, gname=Romeel, sname=Davé]{Romeel Davé}
\affiliation{Institute for Astronomy, University of Edinburgh, Royal Observatory, Edinburgh EH9 3HJ, United Kingdom}
\affiliation{Department of Physics and Astronomy, University of the Western Cape, Robert Sobukwe Rd, Cape Town 7535, South Africa}
\email{Romeel.Dave@ed.ac.uk}

\author[orcid=0000-0001-8156-6281, gname=Tyler, sname=McCabe]{Tyler McCabe}

\affiliation{School of Earth and Space Exploration, Arizona State University, 781 Terrace Mall, Tempe, AZ 85287, USA}
\email{tjmccab2@asu.edu}

\begin{abstract}
We compare O\,\textsc{vi} absorption in synthetic spectra from galaxy groups in the SIMBA and TNG50 cosmological hydrodynamic simulations against those observed from the COS-IGrM survey. We select 14 galaxy groups from each simulation with $12.89 \le \log(M_{\rm halo}/M_\odot) \le 13.61$, closely matching COS-IGrM, and create 90{,}000 synthetic spectra per group. We demonstrate the utility of synthetic absorption spectroscopy when comparing simulations with QSO absorption-based observations. We investigate absorber properties such as radial distributions and kinematics with respect to the group and nearest galaxy. The O\,\textsc{vi} covering fraction ($f_{\rm OVI}$) in TNG50 ($20.62 \pm 2.56\%$) and SIMBA ($5.98 \pm 0.82\%$) are both systematically lower than COS-IGrM ($44 \pm 5\%$). Kinematic analysis reveals that vast majority ($\sim 95\%$) of absorbers in both the simulations are gravitationally bound. In TNG50, strong absorbers ($\log N_{\rm OVI} > 15$) are located near star-forming galaxies ($\log {\rm sSFR} > -11$) within $\sim 200$~kpc, suggesting physical connection to stellar feedback, whereas SIMBA shows no comparable trend. Furthermore, in TNG50 occurrence of O\,\textsc{vi} absorbers at small impact parameters increases with stellar mass of nearest galaxy, but shows no dependence on total stellar mass of group. In contrast, SIMBA shows no clear correlation with nearest galaxy’s stellar mass, though groups with higher total stellar mass exhibit higher detection rate at larger impact parameters. Differences observed in simulations may arise from feedback models and resolution effects. Finally, we show absorber analysis methodology is important factor when comparing simulations with absorption spectroscopy observations.
\end{abstract}

\section{Introduction}  

Most galaxies lie in groups of galaxies ($\sim60\%$) in the Local Universe \citep[e.g.,][]{Tully_1987, Ramella_2002, Eke_2004, Tago_2008}. Galaxy groups are gravitationally bound systems that typically contain a few to a few tens of galaxies and bridge the mass between isolated galaxies and massive galaxy clusters \citep{Tully_1987}. Additionally, galaxy properties tend to change due to environmental effects at the densities found in groups, suggesting that important reprocessing mechanisms operate before groups eventually merge into clusters \citep{Lewis_2002, Gomez_2003}. Galaxy groups also host a significant fraction of cosmic baryons, thus playing a critical role in cosmic structure formation and galaxy evolution \citep{Fukugita_2006, spergel_2007, Tumlinson_2017}.

The diffuse gas gravitationally bound within galaxy groups, known as the intragroup medium (IGrM), plays a critical role in galaxy evolution by regulating key processes such as star formation, galaxy interactions, and chemical enrichment driven by galactic outflows and active galactic nuclei (AGN) feedback \citep{mulchaey_2000, stocke_2019}. The IGrM spans a broad temperature range, typically dominated by a hotter component ($T \gtrsim 10^{6}$~K) that emits primarily in soft X-rays and can also be traced through absorption by highly ionized species such as O\,\textsc{vii} and O\,\textsc{viii} \citep{mulchaey_2000, helsdon_2000, Bregman_2007, Pratt_2020, stocke_2014}. Additionally, a significant fraction of baryons is predicted to reside in a cooler, ``warm-hot" phase with temperatures of $T\sim10^{5}$--$10^{6}$~K, which is challenging to detect directly in emission due to its diffuse nature and lower emissivity. Instead, this cooler gas can be observationally detected through ultraviolet absorption lines, particularly from ions like O\,\textsc{vi}, as well as broad Ly$\alpha$ (BLA) features \citep{stocke_2014, Danforth_2008, Tumlinson_2017}. Even cooler, photoionized gas at temperatures of $T \sim 10^{4}$~K may also exist, further highlighting the complexity and multi-phase structure of the IGrM \citep{Savage_2014}. Accurate characterization of various  phases of the IGrM is essential for a complete census of baryons in galaxy groups and a comprehensive understanding of the IGrM \citep{stocke_2014, Pachat_pair_2016}.

Ultraviolet (UV) absorption spectroscopy toward background quasars (QSOs) offers a robust observational method for detecting diffuse gas in the intermediate-temperature range ($T\sim10^{5}$--$10^{6}$~K), a regime too cool for strong X-ray emission and too hot for neutral gas diagnostics \citep{Rauch_1998, Shull_2012, Savage_2014}. Among the available UV tracers, the O\,\textsc{vi} doublet (1031.93, 1037.62~\AA) is especially valuable: it reaches its peak ionization fraction around $T\approx10^{5.4}$~K with tail down to $T\approx10^{6}$~K under collisional ionization equilibrium, aligning closely with the virial temperatures of galaxy groups \citep{Gnat_2007,Mulchaey_1996, Savage_2014, tumlinson_2013, Werk_2014}. While hotter phases ($T>10^{6}$~K) can be probed by ions such as O\,\textsc{vii}, O\,\textsc{viii}, Ne\,\textsc{viii}, and Mg\,\textsc{x}, these are often difficult to detect due to weaker transitions, lower cosmic abundances, and current instrumental limitations in X-ray and high-SNR UV spectroscopy \citep{Bregman_2007, Werk_2014}. Conversely, ions like C\,\textsc{ii}, Si\,\textsc{ii}, Mg\,\textsc{ii}, C\,\textsc{iii}, and Si\,\textsc{iii} only trace cooler ($T\sim3\times10^{4}$~K), denser gas. Thus, O\,\textsc{vi} absorption remains one of the most accessible diagnostics for studying the warm-hot phase of the intragroup medium.

The interpretation of O\,\textsc{vi} absorbers is complicated due to their potential formation under multiple ionization conditions. Numerical simulations and theoretical studies differ on whether O\,\textsc{vi} arises predominantly through photoionization equilibrium (PIE) at low densities ($n_{\rm H}\sim10^{-5}$~cm$^{-3}$) and cooler temperatures ($T\sim10^{4}$~K) or via collisionally ionized gas at warmer temperatures ($T\sim2-5\times10^{5}$~K) \citep{Kang_2005,Oppenheimer_2009,Oppenheimer_2011,Smith_2011,Tepper_2012,McQuinn_2018}. Despite this ambiguity, O\,\textsc{vi} remains advantageous due to its relatively high cosmic abundance, strong UV absorption transitions, and ionization properties that can be effectively leveraged to distinguish between these competing scenarios through careful measurements of gas temperature and kinematics.

Previous observational efforts with the Cosmic Origins Spectrograph (COS) \citep{Green_2012} aboard the \textit{Hubble Space Telescope}, along with other UV QSO background absorption studies, have greatly expanded our understanding of O\,\textsc{vi} absorption and other species in galaxy groups \citep[e.g.,][]{Tripp_2000a,Tripp_2000b, Savage_2010,Savage_2014,Tripp_2008, Narayanan_2018,stocke_2014,stocke_2017, pointon_impact_2017, stocke_2019,mccabe_2021}. Although these observations reveal abundant O\,\textsc{vi}-bearing gas in group environments, the precise origin of these O\,\textsc{vi}-bearing clouds, their physical association with the group halo or galaxy members, and their properties remain only partially understood.

Cosmological hydrodynamic simulations have developed a physically consistent framework linking the incidence of O\,\textsc{vi} absorption to halo mass, thermodynamic structure, and feedback history. Studies in large cosmological volumes and zoom-in simulations demonstrate that O\,\textsc{vi} is not a simple tracer of total gas mass, but instead arises from a restricted region of temperature--density phase space that depends sensitively on the balance between photoionization and collisional ionization, as well as on the adopted stellar and AGN feedback prescriptions \citep[e.g.,][]{Hummels_2013,Ford_2016,Liang_2016,Suresh_2017,Gutcke_2017,Roca_2019,Marra_2021,Oppenheimer_2021}. In IllustrisTNG, \citet{Nelson_2018_OVI} showed that O\,\textsc{vi} columns peak in
$L_\star$ haloes and decline at higher
masses, where the hotter gas shifts the oxygen budget toward O\,\textsc{vii} and
O\,\textsc{viii}. In EAGLE simulations, \citet{Oppenheimer_2016} and \citet{Wijers_2020} further demonstrate that low-redshift O\,\textsc{vi} exhibits a bimodal origin, with contributions from cooler, low-density photoionized gas and warmer, collisionally ionized plasma. The relative importance of these channels shifts systematically with halo mass, star-formation activity, and environment \citep[see also][]{Stern_2018,Strawn_2021}.

Within galaxy groups, simulations predict a complex and multiphase intragroup medium in which O\,\textsc{vi} traces a transitional regime between hot X-ray emitting gas and cooler photoionized structures, shaped by the cumulative impact of baryonic feedback on the thermodynamic and chemical state of the halo \citep[e.g.][]{Angelinelli_2023,Ayromlou_2023}. \citet{Ayromlou_2023} show that feedback reshapes the baryon distribution in a
mass dependent way, with quenched group haloes exhibiting significantly larger
closure radii than star-forming systems at the same mass-consistent with the
stronger influence of AGN feedback in the group-halo regime
($M_{\rm halo}\!\sim\!10^{12}$--$10^{14}\,{\rm M_\odot}$). Studies also show
that AGN-driven outflows are most effective in group-scale haloes, where feedback energy can
exceed the binding energy of the halo gas, lowering gas fractions and expelling
metal-enriched material into the outer halo
\citep{Davies_2019,Davies_2020,Ayromlou_2023,Angelinelli_2023}. Other simulations and semi-analytic models identify group scale haloes as the regime where non-gravitational physics most strongly perturbs gas density, entropy, and metallicity distributions relative to purely gravitational expectations \citep[e.g.,][]{Oppenheimer_2021,Voit_2020,Sorini_2024,Zinger_2020,Truong_2020,Terrazas_2020}. In this feedback regulated environment, the survival and observability of
O\,\textsc{vi}-bearing gas depend not only on halo mass, but also on the cumulative
feedback history and the coupling between AGN-driven outflows and the surrounding
IGrM. Studies of group scale systems, both in simulations and in
X-ray observations show that AGN activity can generate multiphase outflows,
cavities, shocks, and turbulence that heat, uplift, and mix the intragroup gas,
reshaping its thermal structure and metal distribution
\citep[e.g.,][]{Mccarthy_2010,Angles_2017,Mcdonald_2019,Eckert_2021,Aguerri_2021,
Gastaldello_2021,Hlavacek_2022,Oppenheimer_2021,Cui_2024}.

In this paper, we conduct a systematic comparison between O\,\textsc{vi} absorption properties observed in the COS-IGrM survey \citep{mccabe_2021} and predictions from two publicly available hydrodynamic simulation suites: IllustrisTNG  \citep{Pillepich_2018, Springel_2018, Nelson_2018, Naiman_2018,Marinacci_2018} and SIMBA \citep{Dave_2019}. We investigate the spatial distribution, kinematics, and physical conditions of the warm-hot gas associated with galaxy groups, in order to quantify the role of O\,\textsc{vi} absorbers in the context of the intragroup medium. 

This paper is organized as follows: Section~\ref{sec:methods} outlines the methodologies and the synthetic data used in this study, Section~\ref{sec:results} presents the findings of our comparative analyses. In Section~\ref{sec:discussion}, we present the implication of our results and Section~\ref{sec:conclusion} summarizes our main conclusions and their broader implications for galaxy evolution in group environments.

\begin{figure*}
    \centering
    \includegraphics[width=1\linewidth]{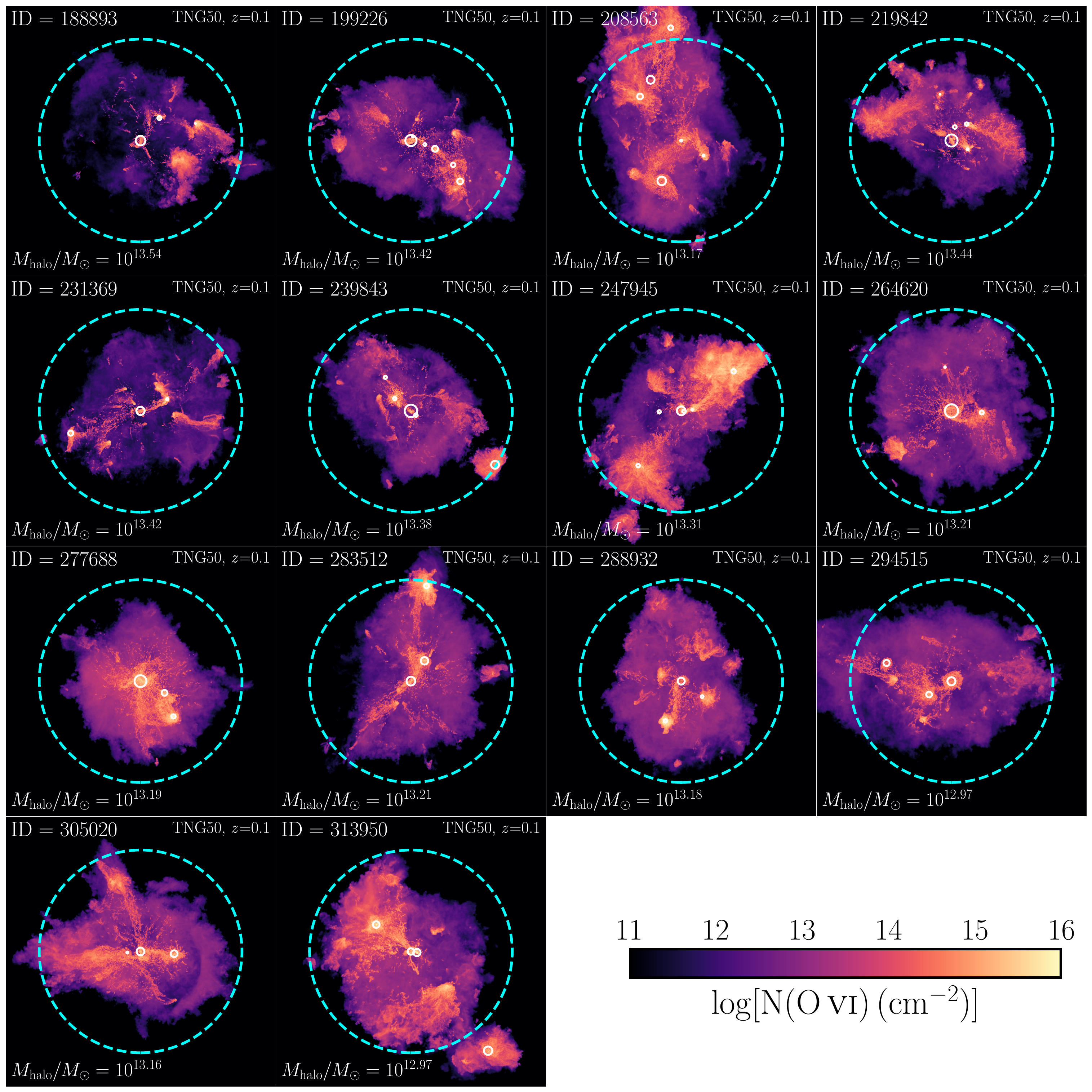}
    \caption{Projected O\,\textsc{vi} column density maps for all selected groups in TNG50 at \( z = 0.1 \), as listed in Table~\ref{tng_group_table}. In each panel, the dashed cyan circle has a radius of \( 1.5R_{200c} \), while the box spans \( 4R_{200c} \) along both axes. All synthetic sightlines are distributed within the region enclosed by the cyan dashed circle and share the same orientation as the native x--y plane of the simulation. The ID of the central subhalo of each group is indicated in the top-left corner, and the halo mass is shown in the bottom-left corner. Smaller white circles represent member subhalos with \( L \geq L^* \), where their radii are three times the stellar half-mass radius. All halos exhibit a concentration of O\,\textsc{vi} reservoirs near the group center. Groups such as ID = 208563, 247945, 283512, and 313950 also have significant O\,\textsc{vi} distributed outside the \( 1.5R_{200c} \) region. \textit{Note: the maps include only gas particles within the friends-of-friends (FoF) halo boundaries; the regions appearing black correspond to areas outside the FoF scope.}}
    \label{fig:projection_tng}
\end{figure*}

\section{Methods} \label{sec:methods}

\subsection{Simulations: IllustrisTNG and SIMBA}

We select simulations with sufficiently large volumes to provide statistically robust and representative sample of galaxy groups while also having the necessary mass and spatial resolution to resolve O\,\textsc{vi}, which can exist in a diffuse state. In particular, we use the TNG50-1 \citep{Nelson_2019_TNG50,Pillepich_2019_TNG50} (hereafter TNG50) run of the IllustrisTNG simulation suite, which has a box size of \(\approx 50\) Mpc.  It has a gas cell mass resolution of approximately \(m_{\mathrm{gas}} \approx 8.5 \times 10^{4} M_{\odot}\) and a dark matter mass resolution of \(m_{\mathrm{dm}} \approx 4.5 \times 10^{5} M_{\odot}\). TNG50 is the highest spatial and mass resolution simulation available at this volume and is also the largest-volume cosmological simulation at this resolution.

For comparison, we also use the large volume \(100 \, h^{-1} \, \mathrm{Mpc}\) cosmological run of the SIMBA simulation suite \citep{Dave_2019}. It has a mass resolution of \(1.82 \times 10^7 \, M_\odot\) for gas cells and \(9.6 \times 10^7 \, M_\odot\) for dark matter particles. In this work, we adopt the version of SIMBA that has been re-processed to enable an apples-to-apples comparison with TNG50. As a result, in both simulations the identification of structures like the halos (groups), subhalos (galaxies), and their properties are identified using the friends of friends (\texttt{FOF}) and \texttt{SUBFIND} algorithms enabling identical analysis routines. \citep{Springel_2001,Dolag_2009, Davis_1985,More_2011}.

The two simulations differ in their treatment of feedback processes, which directly influence ionization conditions and metal distribution in group environments.
IllustrisTNG employs a dual-mode AGN feedback model, where black holes at high accretion rates release thermal energy into the surrounding gas (quasar mode). At low accretion rates, the feedback transitions into a kinetic mode, stochastically injecting momentum into the CGM, driving large-scale winds that heat and redistribute gas in group environments \citep{10.1093/mnras/stw2944, 2017MNRAS.465.3291W, 10.1093/mnras/stx3040, 10.1093/mnras/stx3112}.

On the other hand, SIMBA implements a kinetic AGN feedback model across all accretion states, where gas is preferentially ejected in a bipolar manner, aligned with the angular momentum of the inner accretion disk \citep{Dave_2019}. At low accretion rates, X-ray feedback is introduced, further heating and expelling gas from the CGM, leading to efficient metal ejection and quenching of massive galaxies \citep{Dave_2019}. Additionally, SIMBA adopts a torque-limited black hole accretion model in which gas inflow driven by gravitational torques regulates SMBH growth by coupling accretion to host-galaxy properties, allowing black holes and galaxies to evolve toward observed scaling relations without requiring strong, explicit self-regulation via feedback loops \citep{Angles_2017}. This accretion prescription, together with SIMBA's AGN feedback implementation, is expected to influence the thermal and ionization conditions of the IGrM.

\subsection{Group Selection} \label{subsec:group_selection}

For a direct comparison between the simulations and the COS-IGrM survey \citep{mccabe_2021}, we adopt a similar selection criterion for the simulated group sample, enabling a consistent and physically meaningful comparison between observed and synthetic sightlines. We select the simulation snapshot to match the median redshift of the COS-IGrM survey, \( z = 0.134 \). While for SIMBA we use the snapshot at that redshift, for TNG50-1, we use the closest full snapshot at \( z = 0.1 \), which contains all the necessary fields, instead of the closest mini snapshot at \( z = 0.13 \), which only includes a subset of all particle fields.
The halo masses of the groups were chosen to be \(
 10^{12.89} M_\odot \leq M_{\mathrm{halo}} \leq 10^{13.61} \, M_\odot
\) which corresponds to the range of the COS-IGrM sample. Additionally, groups were selected for this study only if they contained between 3 and 7 subhalos (galaxies) with equivalent SDSS \( r \)-band luminosities of \( L \geq L^* \). These criteria ensured that the sample selection closely matched the selection criteria used by \citet{mccabe_2021} for the COS-IGrM survey. We identify 14 groups in TNG50 (Table \ref{tng_group_table}) based on the filtering criteria above and 182 groups in SIMBA. To maintain consistency in the analysis, we select 14 out of 182 groups in SIMBA with a halo mass distribution similar to the TNG50 sample (Table \ref{simba_group_table}). Fig. \ref{fig:projection_tng} shows the O\,\textsc{vi} projected column density for all the selected groups in TNG50 used in this study. 

We create 90,000 synthetic sightlines and their synthetic spectra per group for both simulations, resulting in a total of \(1.26 \times 10^6\) rays for each simulation. All sightlines for a halo were chosen with a random position within a uniformly binned distribution in $\rho$ and $\theta$, where $\rho$ is the impact parameter and $\theta$ is the angle from the axes in the plane perpendicular to the sightline.
We limit the maximum impact parameter for a sightline to \(1.5R_{\mathrm{vir}}\), and all sightlines traverse from \(-z\) to \(+z\) direction of the native simulation volume coordinates. The total traced path length is also fixed at \(4R_{\mathrm{vir}}\), as the rays extend from \(-2R_{\mathrm{vir}}\) to \(+2R_{\mathrm{vir}}\) relative to the group center along the \(z\)-direction. We adopt this path length to isolate gas physically associated with the group halo, consistent with the isolated-group selection in the COS-IGrM survey. We find that the incremental O\,\textsc{vi} contribution to the integrated column density and covering fraction from path lengths beyond $\pm 2\,R_{\mathrm{vir}}$ drops significantly. These changes are within the uncertainities quoted in the observations and our measurements. This is highlighted in Figure~\ref{fig:full_panel_path} in the Appendix. To generate the synthetic absorption spectra for O\,\textsc{vi}, we use the total oxygen mass per gas cell tracked in the simulations. Its ionization states, accounting for both photo and collisional ionization are computed with \texttt{CLOUDY} \citep{ferland20132013releasecloudy} using the ultraviolet background model of \citet{UVB_2009} (2011 update). These spectra are part of SALSA, the Synthetic Absorption Line Spectral Almanac \citep{nelson2025_salsa} where further implementation details are provided. We then create a mock sightline and use the gas distribution of the ray comprising cells to generate the synthetic spectra having COS-G130M resolution element of 0.01\AA. The resulting spectra are then binned to 3 pixels (half the resolution element) and apply gaussian noise with an SNR=10 per resolution element to match the observational data in \citet{mccabe_2021}.
 
We remove all the sightlines within three times the stellar half-mass radius \(R_{\mathrm{1/2,\star}}\) of the nearest group galaxy, i.e. \(\rho_{cg} \leq 3\,R_{\mathrm{1/2,\star}}\) in the group with \( L \geq L^* \). In the cases where a single sightline contains multiple absorbers, we combine them following the method in \cite{mccabe_2021}.

\begin{table*}
\centering
\caption{TNG50 and SIMBA Selected Groups\label{tab:combined}}
%
%
\begin{subtable}[t]{0.45\textwidth}
\centering
\caption{TNG50-1\label{tng_group_table}}
\begin{tabular}{ccccc}
\hline
ID & $\log[M_{c200}]$ & $R_{c200}$ & $N_{total}$ & $N_{filtered}$ \\
\hline
188893 & 13.54 & 494.34 & 10333 & 3 \\
199226 & 13.42 & 453.76 & 9337  & 6 \\
208563 & 13.17 & 373.52 & 11279 & 6 \\
219842 & 13.44 & 457.87 & 11527 & 6 \\
231369 & 13.42 & 452.03 & 8474  & 3 \\
239843 & 13.38 & 440.11 & 8102  & 5 \\
247945 & 13.31 & 416.73 & 9360  & 6 \\
264620 & 13.21 & 384.19 & 7608  & 3 \\
277688 & 13.19 & 379.33 & 5824  & 3 \\
283512 & 13.21 & 383.72 & 5420  & 3 \\
288932 & 13.18 & 377.35 & 5583  & 3 \\
294515 & 12.97 & 319.22 & 6020  & 4 \\
305020 & 13.16 & 371.58 & 4853  & 3 \\
313950 & 12.97 & 320.57 & 4113  & 4 \\
\hline
\end{tabular}
\end{subtable}
\quad
%
%
\begin{subtable}[t]{0.45\textwidth}
\centering
\caption{SIMBA\label{simba_group_table}}
\begin{tabular}{ccccc}
\hline
ID & $\log[M_{c200}]$ & $R_{c200}$ & $N_{total}$ & $N_{filtered}$ \\
\hline
26911  & 13.54 & 508.83 & 341  & 6 \\
36393  & 13.45 & 471.97 & 198  & 7 \\
38027  & 13.40 & 455.82 & 173  & 6 \\
38659  & 13.39 & 450.70 & 190  & 5 \\
41684  & 13.38 & 448.38 & 142  & 7 \\
45914  & 13.32 & 426.63 & 73   & 4 \\
46543  & 13.19 & 386.48 & 127  & 5 \\
46997  & 13.29 & 417.61 & 117  & 3 \\
48262  & 13.22 & 396.71 & 106  & 4 \\
51199  & 13.15 & 376.41 & 113  & 3 \\
52263  & 13.19 & 388.12 & 90   & 3 \\
53374  & 13.19 & 388.64 & 77   & 4 \\
57091  & 12.96 & 324.53 & 62   & 4 \\
69625  & 12.98 & 330.98 & 58   & 4 \\
\hline
\end{tabular}
\end{subtable}
%
%
\tablecomments{\emph{ID} refers to the central subhalo ID;
$M_{c200}$ is the virial mass in \(M_{\odot}\); 
$R_{c200}$ is the virial radius in \(ckpc/h\); 
$N_{total}$ is the total number of subhalos in the group;
$N_{\mathrm{filtered}}$ is the number of subhalos that satisfy the COS-IGrM selection criteria, i.e., groups that contain $3 \leq N_{L \geq L^*} \leq 7$ subhalos (galaxies) where L is the SDSS $r$-band equivalent luminosity. 
}
\end{table*}

\subsection{Spectral fitting}\label{subsec:spectral_fittting}
We use the publicly available python package \texttt{PYGAD} \citep{2020MNRAS.496..152R} for Voigt profile fitting for all the synthetic spectra, which employs non-linear least-squares minimization. The fitting procedure systematically identifies absorption regions and iteratively adds absorption components until the reduced chi-squared value reaches the provided threshold or the maximum number of components allowed. In our pipeline, we set the chi-squared threshold to \( \chi^2 = 1.0 \) and limit the number of lines per region to a maximum of 6 to prevent overfitting. The best-fit parameters include the equivalent width \(W_{\lambda}\), centroid wavelength \( \lambda \), Doppler parameter  \( b \), column density \( N \), and their uncertainties, which are derived from the covariance matrix of the optimization process \citep{2020SciPy-NMeth}. To prevent false detections and overfitting, a minimum region width of 3 pixels is chosen, and the detection significance threshold is set to \( 3\sigma \) above the noise level. Absorption features within \( \pm800 \) km/s of the systemic velocity \( v_{\text{sys}} \), corresponding to the simulation group's redshift, are identified. Absorbers beyond \( \pm800 \) km/s are considered to be dominated by the Hubble flow and not associated with the group, a criterion similar to observations, and are therefore excluded from the analysis. In cases where no absorption region is identified, we estimate upper limits by computing the limiting equivalent width \( W_{3\sigma} \) and the corresponding upper limit column density \( \log(N_{3\sigma}) \) within 100$\text{ km s}^{-1}$ range. We adopt a \( 3\sigma \) threshold for this calculation, resulting in an upper limit column density of \( \log(N_{3\sigma})\) of 13.49.  To maintain physically meaningful fits, we constrain \( \log[N(\text{O\,\textsc{vi}})] \) in the range 13.49 to 18.0 . We also impose a reasonable constraint on the Doppler parameter \( b \), requiring \( 6\,\text{km s}^{-1} \leq b \leq 100\,\text{km s}^{-1} \).

\section{Results}\label{sec:results}

\subsection{Validation of Absorber Recovery Against Observational Effects}\label{subsec:obs_effects_and_bias}

\begin{figure*}[t]
    \centering

    \begin{subfigure}[t]{0.5\linewidth}
        \centering
        \includegraphics[width=\linewidth]{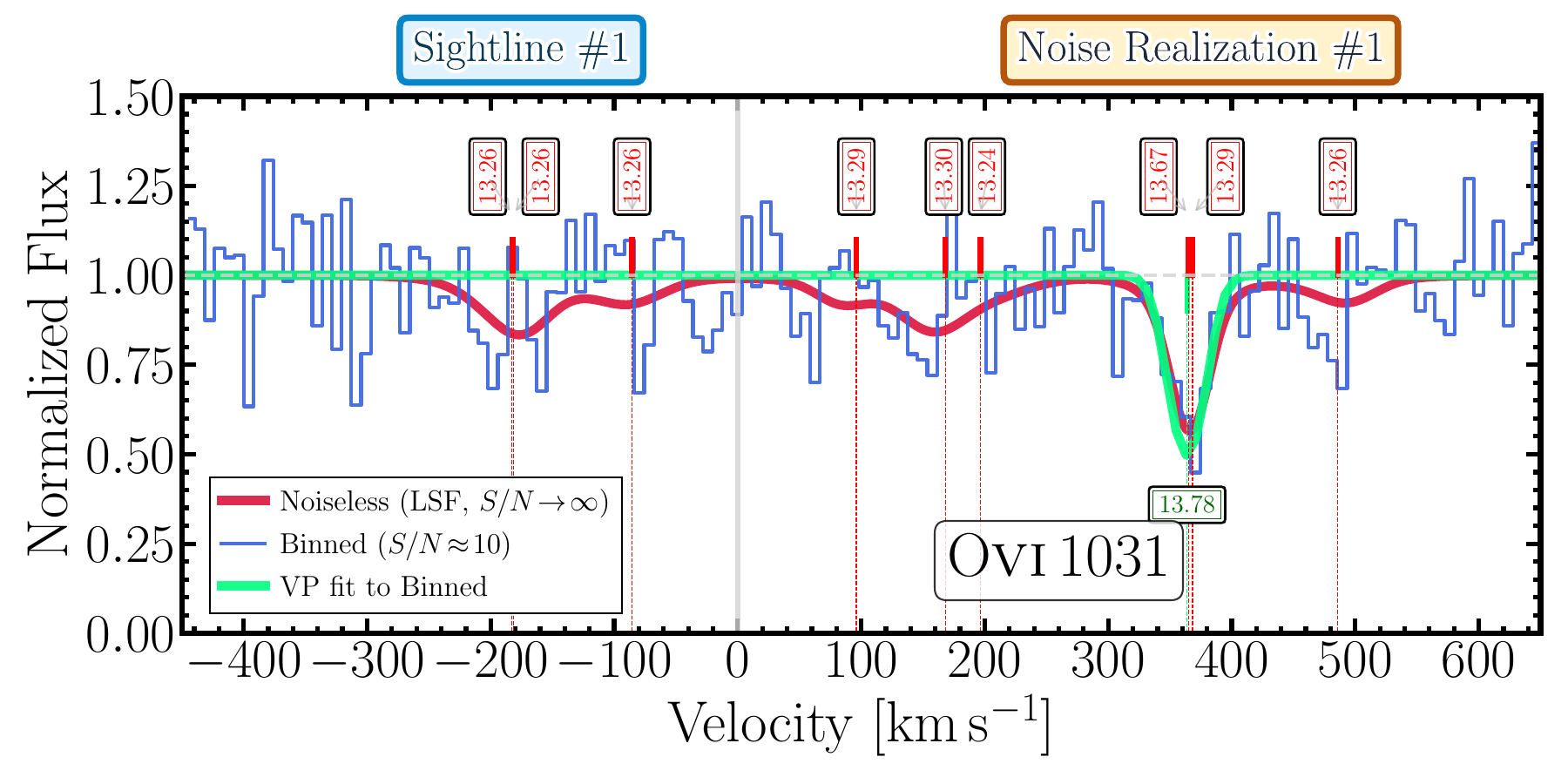}
        \caption*{}
        \label{fig:s1_seed071}
    \end{subfigure}%
    \begin{subfigure}[t]{0.5\linewidth}
        \centering
        \includegraphics[width=\linewidth]{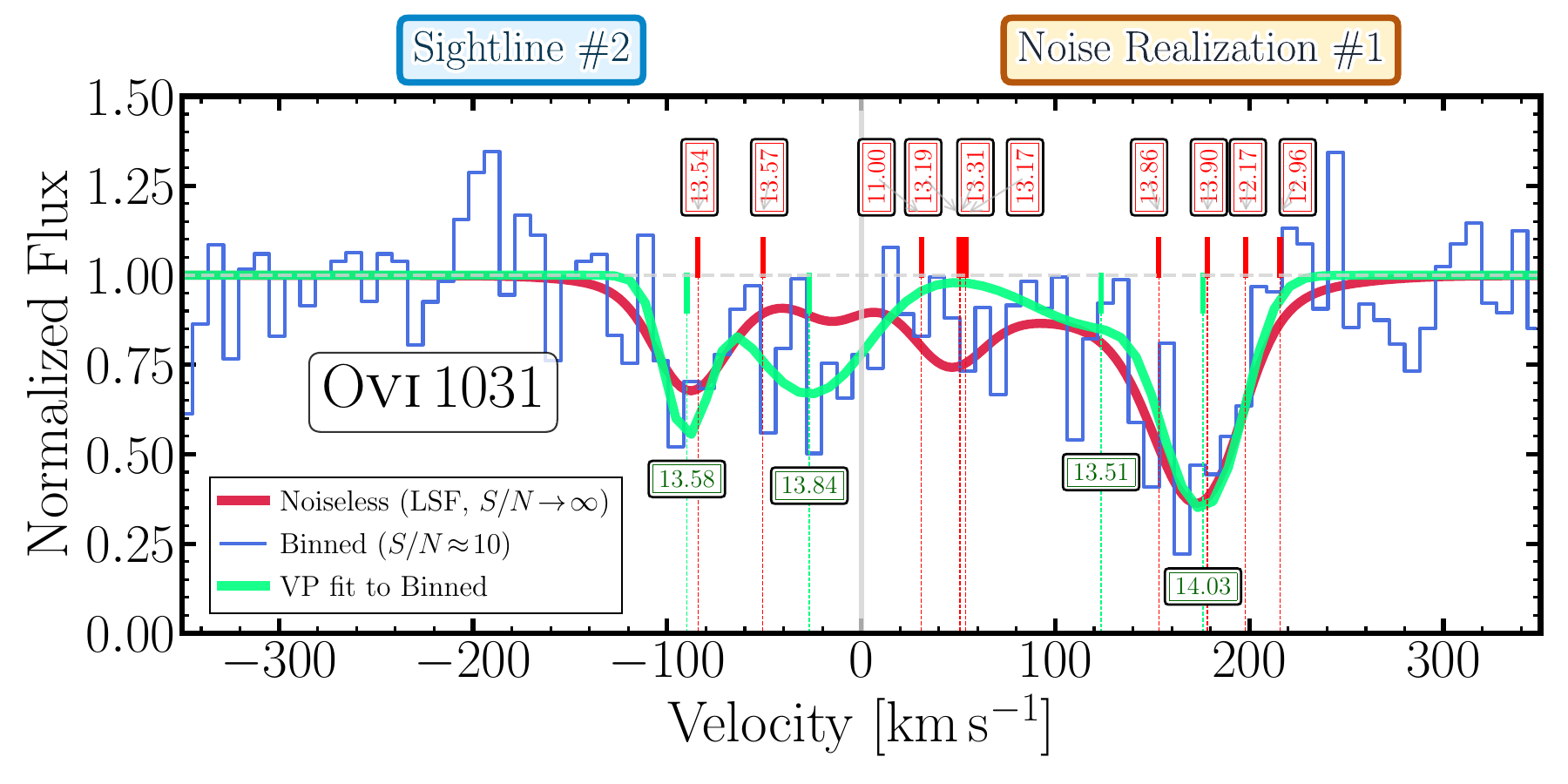}
        \caption*{}
        \label{fig:s2_seed107}
    \end{subfigure}

    \vspace{-7mm}

    \begin{subfigure}[t]{0.5\linewidth}
        \centering
        \includegraphics[width=\linewidth]{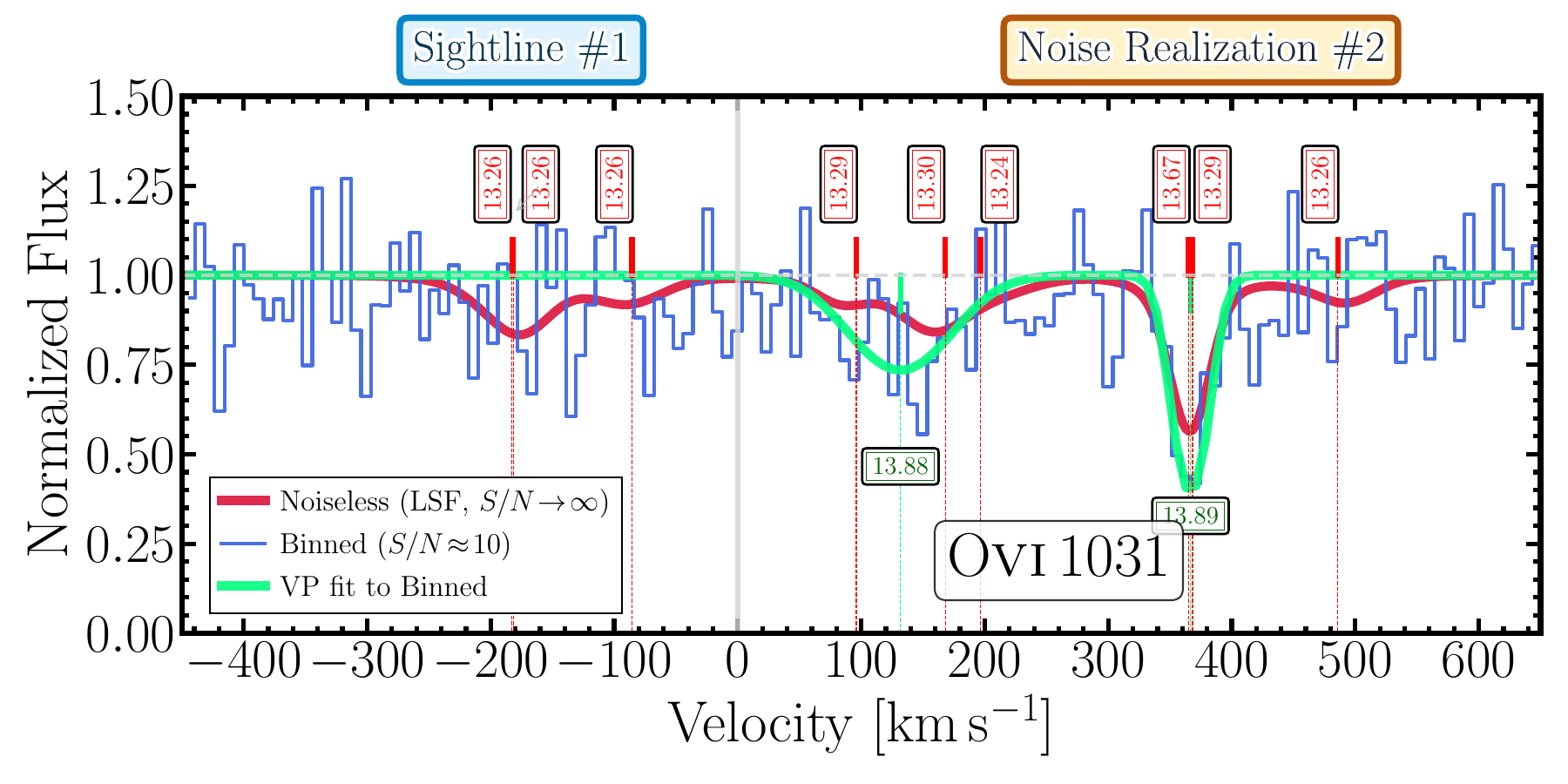}
        \caption*{}
        \label{fig:s1_seed023}
    \end{subfigure}%
    \begin{subfigure}[t]{0.5\linewidth}
        \centering
        \includegraphics[width=\linewidth]{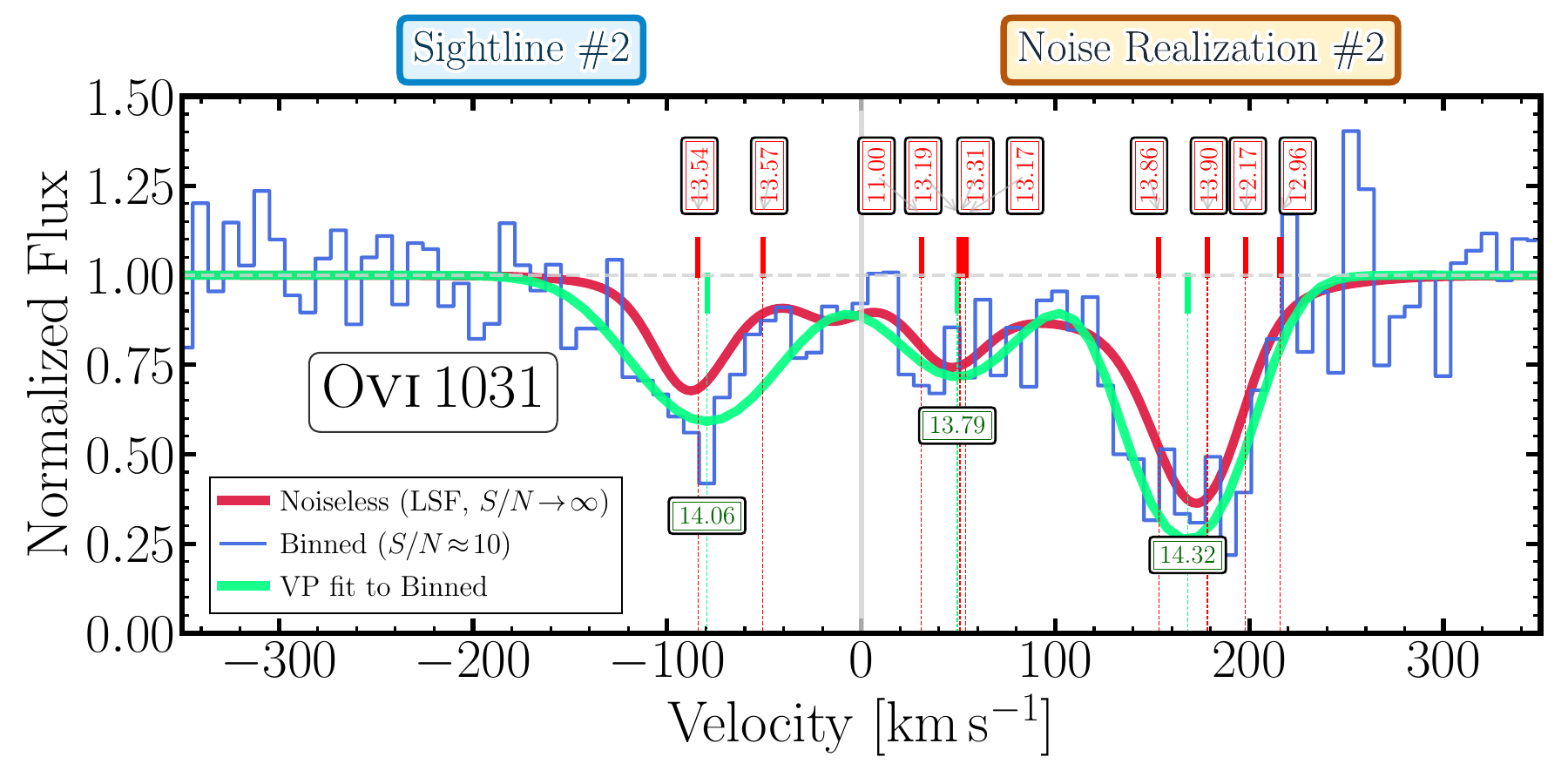}
        \caption*{}
        \label{fig:s2_seed118}
    \end{subfigure}

    \vspace{-7mm}

    \begin{subfigure}[t]{0.5\linewidth}
        \centering
         \includegraphics[width=\linewidth]{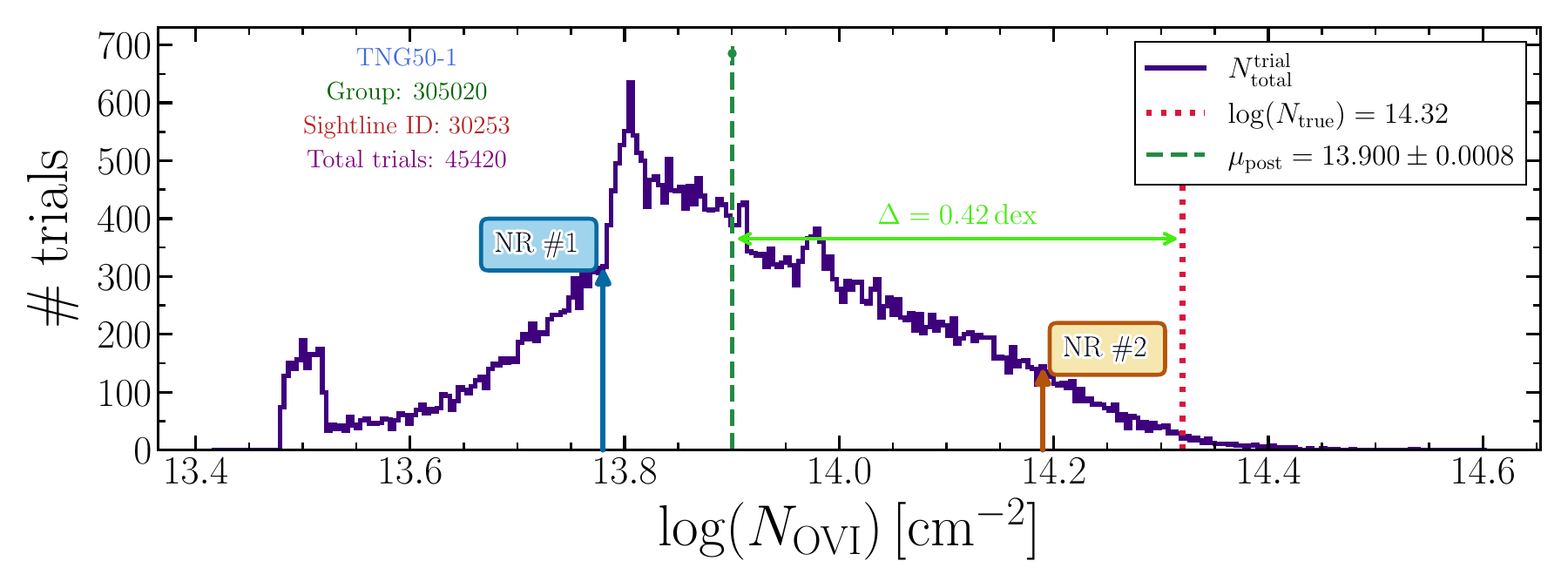}
        \caption*{}
        \label{fig:s1_marginal}
    \end{subfigure}%
    \begin{subfigure}[t]{0.5\linewidth}
        \centering
        \includegraphics[width=\linewidth]{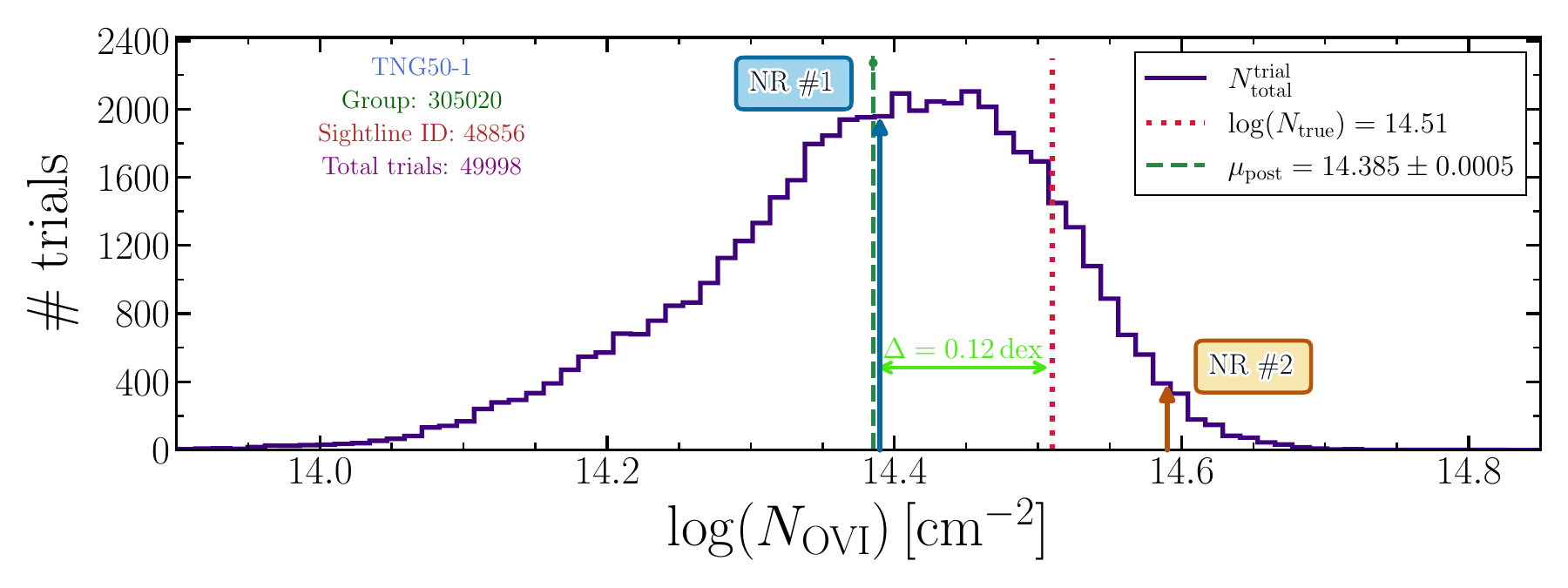}
        \caption*{}
        \label{fig:s2_marginal}
    \end{subfigure}
    \vspace{-7mm}
    \caption{%
    Spectral–posterior diagnostics for two representative TNG50 sightlines in group\,305020.    
    \emph{Rows\,1–2:} True infinite-SNR spectra at COS-G130M resolution convolved with the line-spread function (solid red); the same spectra after binning and addition of noise (blue); and the multi-component Voigt-profile fits to these noisy spectra (solid green). Each fitted or true component is annotated at its velocity centroid by a coloured tick and its column density ($\log N_{\mathrm{OVI}}$).  
    \emph{Row\,3:} Histograms of total column densities from $\sim5\times10^{4}$ independent noise realisations of the same spectra (purple step curves). The vertical dashed lines mark the posterior means $\mu_{\mathrm{post}}$. The dotted crimson lines indicate the true column densities $N_{\mathrm{true}}$ and $N_{\mathrm{total}}$ of individual noise realizations (NR) shown in rows above are annotated by arrows.
    }
    \label{fig:all_spectra_and_marginals}
\end{figure*}

\begin{figure}[htbp]
    \centering
    \includegraphics[width=\linewidth]{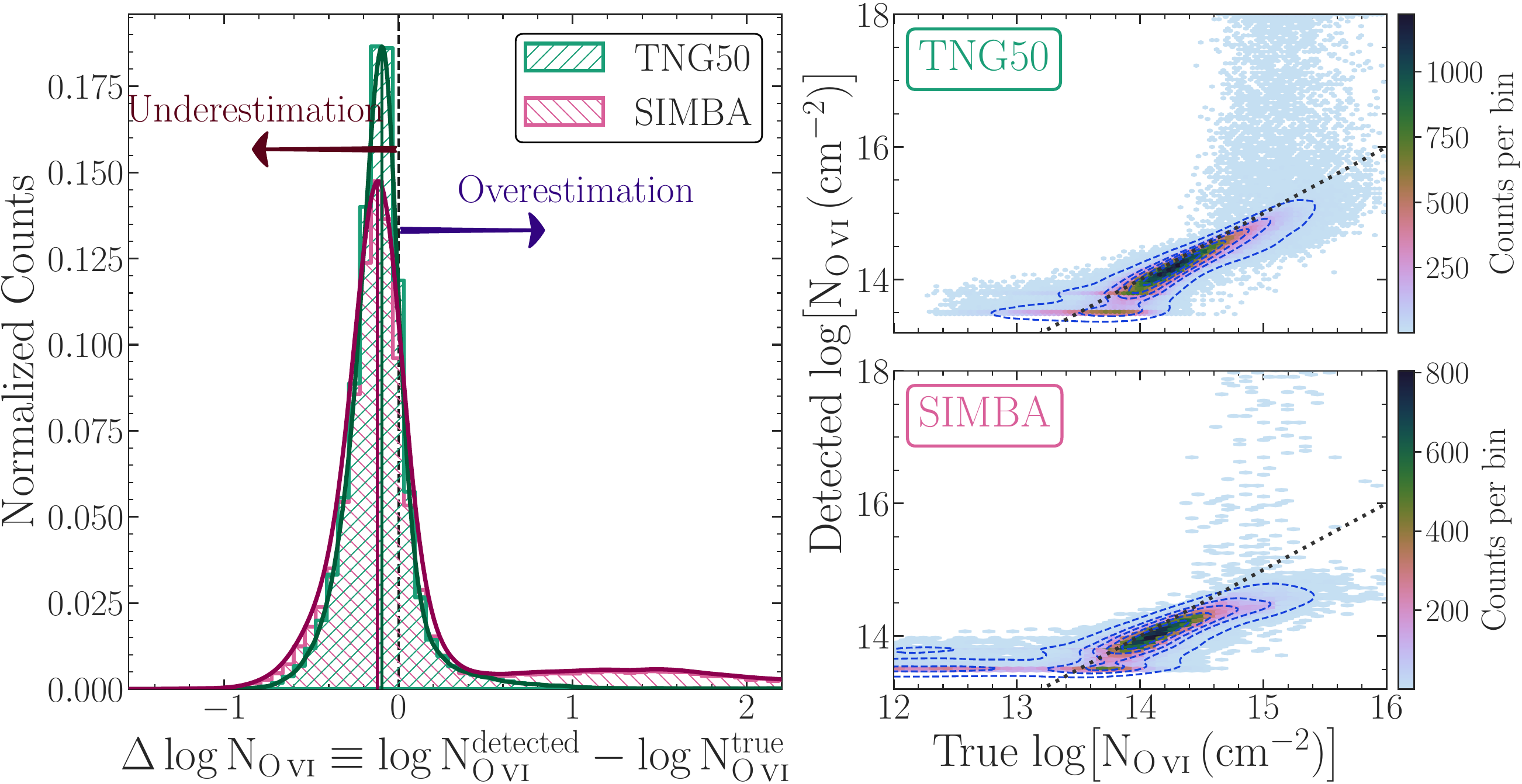}
    \caption{\emph{Left:} Normalized histogram of $\log N_{\rm detected} - \log N_{\rm true}$ showing systematic underestimation and overestimation of absorber strengths in the full sample of detections in TNG50 and SIMBA.
    \emph{Right:} Distribution of detected versus true column densities for TNG50 (top) and SIMBA (bottom). Dotted black line corresponds to the one-to-one relation.}
    \label{fig:delta_logN_tng_simba}
\end{figure}

\begin{figure*}
    \centering
    \includegraphics[width=1\linewidth, trim={50 10 0 10}]{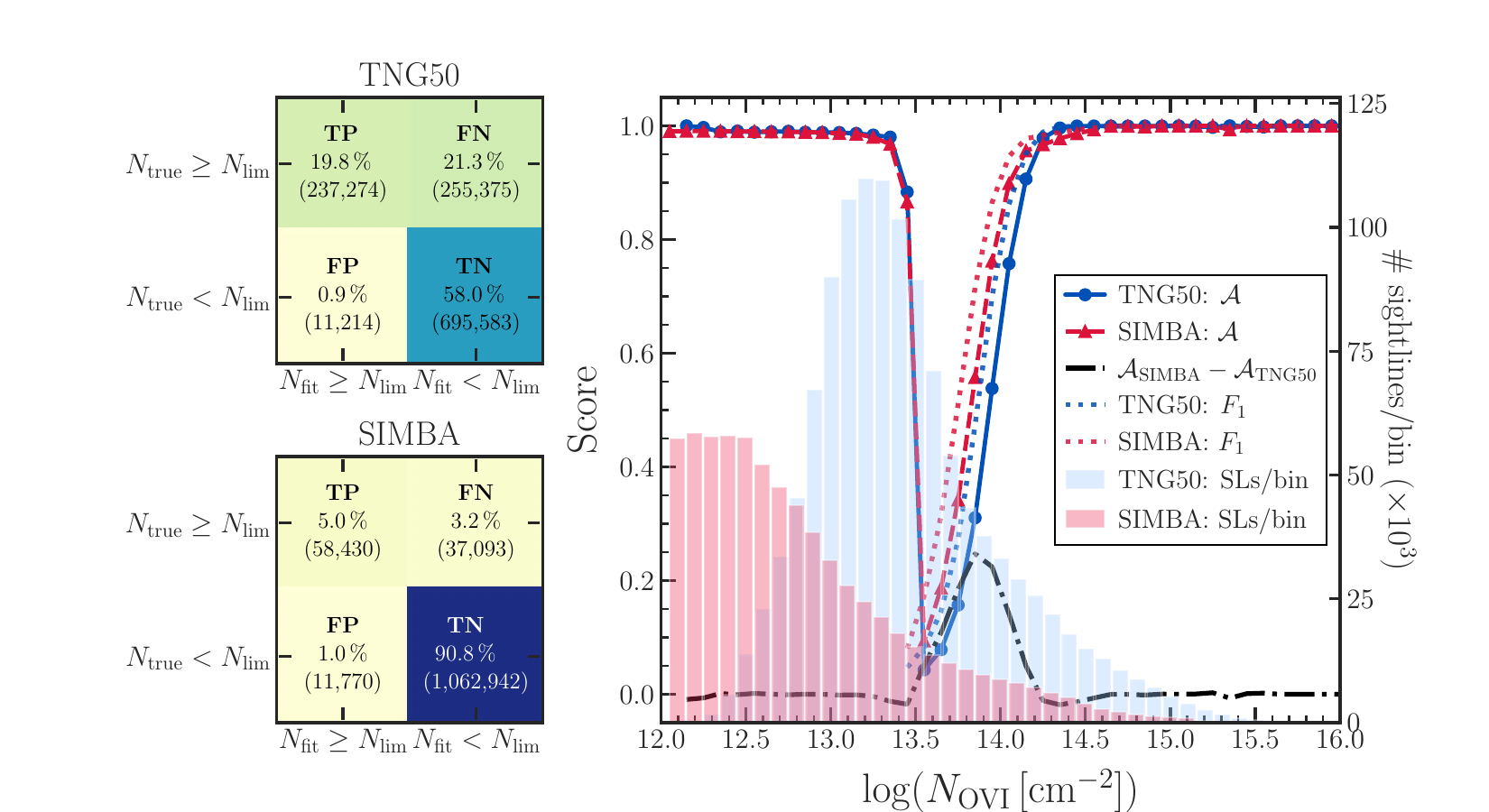}
    \caption{Diagnostic comparison of the \textsc{TNG50} and \textsc{SIMBA} with fitting pipeline.  
\textit{Left:} $2\times2$ confusion matrices for an O\,\textsc{vi}‐column–detection threshold of $N_{\mathrm{lim}} = 13.5$.  Each tile lists the class label (TP, FP, FN, TN), the fraction of all predictions in that cell (colour–coded from 0 to 100\,\%), and the absolute count of sightlines.  
\textit{Right:} Accuracy (\(\mathcal{A}\)) performance as a function of true column density.  Circles and triangles trace the per-bin accuracy for \textsc{TNG50} and \textsc{SIMBA} (0.1-dex bins); dotted lines give the corresponding $F_{1}$ scores; the black dot–dashed curve shows the accuracy difference $\Delta\mathcal{A} = \mathcal{A}_{\text{SIMBA}} - \mathcal{A}_{\text{TNG50}}$.  Coloured bars on the twin axis indicate the number of sightlines per bin.  Together, the panels illustrate that while \textsc{SIMBA} attains a higher global accuracy and $F_{1}$, most of the improvement arises in the low-$N$ regime where \textsc{TNG50} suffers larger false-negative rates.%
}
    \label{fig:confusion_matrix}
\end{figure*}

The necessity of employing the synthetic spectroscopy approach, rather than directly using integrated column density, arises fundamentally from the need to ensure a realistic comparison with observational data \citep[e.g.,][]{Nelson_2018_OVI,2020MNRAS.496..152R}. Direct analysis of integrated column density maps derived from the simulations do not account for the observational effects such as instrumental resolution, sensitivity limitations, and noise characteristics, potentially leading to the over or underestimation of absorber strengths, particularly in the cases where the sightline consists of multiple broad absorbers or the absorbers have the real column density close to the detection limit of the instrument. It needs to be emphasized that this may cause significant discrepancies between the simulated intrinsic absorber strengths and observational measurements. We attempt to observe and quantify such effects that would be applicable to all absorption spectroscopy based surveys.

\subsubsection{Impact of Noise and Resolution}\label{subsubsec:noise_and_res}

An example of the impact of noise and instrument resolution is illustrated in Figure~\ref{fig:all_spectra_and_marginals}, where we present the analysis of the spectra for two representative sightlines from the TNG50 simulation group ID 305020. Focusing initially on first sightline (left), the top two rows compare the true synthetic spectrum shown in the solid red curve, representing infinite signal-to-noise ratio ($\mathrm{SNR}=\infty$) convolved with the COS-G130M instrument line-spread function. When noise is added to get a $S/N =10$, we get the blue spectrum and we show two different noise realizations in the top and middle rows. In the true, noise–free spectrum a total of five individual components are identified. Each component is marked at its velocity centroid with a red tick, and its fitted column density is displayed in the accompanying red label. Summing these five components yields a total fitted O\,\textsc{vi} column density of  
\(
\log N_{\mathrm{OVI}} = 14.32^{+0.032}_{-0.030}
\). This is in excellent agreement with the value obtained directly from the contributing gas cells along the sightline,  
(\(\log N_{\mathrm{OVI,true}} = 14.32\)) (noise realization \#1). Conversely, the solid green curve, representing the multi-component Voigt-profile fit to the noisy spectrum, yields a significantly lower total column density of \(\log N_{\mathrm{OVI}} = 13.78 \pm 0.3\), as only the strongest component is detected while weaker or broader components near the detection limit are missed. This example clearly demonstrates how observational noise and limited resolution can suppress or blend genuine absorption features, leading to a systematic under-estimate of the total column density.

We also perform a similar comparison for second sightline (right column of Figure~\ref{fig:all_spectra_and_marginals}) to highlight other discrepancies. The integrated column density is well recovered (\(\log N_{\mathrm{OVI}} = 14.44^{+0.55}_{-0.44}\)) (noise realization \#1), consistent with the true value (\(\log N_{\mathrm{OVI,true}} = 14.51\); \(\Delta \approx 0.06\) dex). Again we observe the same trend of noise combined with the line-spread function leading to errors in component detection and measurement. For example, the component at \(-90~\mathrm{km\,s^{-1}}\) is enhanced and detected as a feature, while the component at \(59~\mathrm{km\,s^{-1}}\) is suppressed and not detected. Such discrepancies not only affect column density measurements, they can also significantly bias kinematic information. Noise Realizations \#2 for both the sightlines (middle row in Figure~\ref{fig:all_spectra_and_marginals} ) further demonstrates the cases where noise amplifies multiple components, leading to an overestimation of their individual detected strengths.

\subsubsection{Population-Level Bias in Absorber Detection}

The impact of noise can be assessed more broadly by repeatedly adding random noise to the same spectrum and re-measuring the total column density. For each processed spectrum of examples in \ref{subsubsec:noise_and_res} we generated $5\times10^{4}$ independent noise realizations at fixed $S/N$. This produces a distribution of recovered \(N_{O_{VI}}\) values, shown in the bottom row of Figure~\ref{fig:all_spectra_and_marginals}. The vertical dashed line marks the posterior mean estimate of this distribution ($\mu_{\mathrm{post}}$), the dotted crimson line shows the true intrinsic column density, and the horizontal arrow highlights the offset $\Delta = \mu_{\mathrm{post}} - \log N_{\mathrm{true}}$.

For first sightline, the posterior mean is $\mu_{\mathrm{post}} \simeq 13.9$, compared to a true value of $\log N_{\mathrm{OVI}} = 14.32$. This corresponds to an underestimation of about $0.4$ dex. In contrast, the second sightline yields $\mu_{\mathrm{post}} \simeq 14.39$, close to its true value of $14.51$, a much smaller bias of only $\sim0.1$ dex.  We observe the trend that when a spectrum is dominated by weak absorption components (first sightline), the total column density is systematically underestimated, because many features lie close to the detection threshold and are easily suppressed by noise. For spectra dominated by strong absorbers (second sightline), the recovery is much more reliable. This demonstrates that bias cannot be treated as global, instrument only constants, but instead depends on the specific combination of absorber strengths, blending, and velocity dispersions. The effect becomes particularly pronounced near the detection limit, whereas strong absorbers remain comparatively unaffected.

\subsubsection{Recovery Statistics Across the Full Sample}

In Figure~\ref{fig:delta_logN_tng_simba}, we illustrate the accuracy of recovered O\,\textsc{vi}  column density from synthetic spectra in \textsc{TNG50} and \textsc{SIMBA}. The left panel shows the normalized distribution of 
$\Delta \log N_{\mathrm{OVI}} \equiv \log N_{\mathrm{detected}} - \log N_{\mathrm{true}}$ for all detected absorbers, capturing both systematic underestimation and overestimation of column densities. In both simulations the distributions peak at $\Delta \log N_{\mathrm{OVI}} \simeq -0.1$ dex, indicating a slight overall bias toward underestimating absorber strength. The right-hand panels compare the detected and intrinsic column densities for individual absorbers in each simulation. The clustering of points along the one-to-one relation demonstrates that relatively strong systems are generally recovered with high fidelity, whereas the increasing scatter toward lower column densities reflects growing measurement uncertainty near the detection threshold ($\log N_{\mathrm{OVI}} \sim 13.5$). We note that similar deviations are also present for some intrinsically strong absorbers, where recovered column densities are overestimated due to effects such as line blending and saturation.

While Fig.~\ref{fig:delta_logN_tng_simba} characterizes systematic biases in the recovered column densities of detected absorbers, the overall fidelity of the detection process is better evaluated through classification statistics across the entire sightline sample. Fig. ~\ref{fig:confusion_matrix} therefore extends the recovery analysis to all synthetic sightlines from both \textsc{TNG50} and \textsc{SIMBA}. Recovery is summarized using a confusion matrix, which records whether each sightline is classified as a true positive when a genuine absorber is correctly detected, a false positive when a spurious detection is reported, a false negative when a genuine absorber is missed, or a true negative when a non-detection is correctly identified. From these outcomes we derive several standard performance metrics. Purity measures the probability that a reported detection is real, completeness measures the fraction of genuine absorbers that are recovered, accuracy ($\mathcal{A}$) records the overall fraction of correct classifications, and the $F_{1}$ score combines purity and completeness through their harmonic mean.  
The right-hand panel of Figure~\ref{fig:confusion_matrix} shows how these metrics vary with absorber strength. Both simulations achieve nearly perfect recovery, with $\mathcal{A} > 0.95$ and $F_{1} \simeq 1$ for strong absorbers with $\log N_{\mathrm{OVI}} \gtrsim 14.2$. Below this threshold the behavior diverges. In \textsc{TNG50}, accuracy declines to about 0.3 and $F_{1}$ falls to 0.5 by $\log N_{\mathrm{OVI}}=13.8$, indicating that many genuine absorbers are lost to noise. In \textsc{SIMBA}, the recovery remains higher at the same column density, with $\mathcal{A} \approx 0.55$ and $F_{1} \approx 0.7$.
When integrated over the full sample, the overall performance corresponds to $\mathcal{A}=0.78$ and $F_{1}=0.63$ for \textsc{TNG50}, compared with $\mathcal{A}=0.96$ and $F_{1}=0.71$ for \textsc{SIMBA}. The apparent superiority of the method when applied to \textsc{SIMBA} does not arise from the method itself but rather from the underlying absorber distribution. In \textsc{TNG50} the distribution peaks at $\log N_{\mathrm{OVI}} \simeq 13.2$, only 0.3 dex above the detection threshold, so a large fraction of absorbers are susceptible to being suppressed by noise. In contrast, most \textsc{SIMBA} absorbers lie further below the detection threshold ($\log N_{\mathrm{OVI}}\simeq12.2$), and therefore more of them are successfully recovered.  
These findings emphasize two important points. First, any spectroscopic observation, irrespective of the ion species considered, is constrained by instrumental sensitivity and noise, so the most immediate question is whether a given absorber can be detected at all. Second, even when detection is possible, the recovered column density can still differ systematically from the intrinsic value. The magnitude of this effect is not determined by the instrument alone but also by the interplay between underlying distribution of absorber strengths in the population and the sensitivity. Systems dominated by strong absorbers are almost always detected and have reliable measurements, whereas weak systems lying close to the detection threshold are often missed, falsely detected or have unreliable measurements. As a result, comparisons between simulations or with observational surveys must carefully account for these differences.
\subsection{Comparison with COS-IGrM}\label{subsec:cos_igrm_comparoson}

\begin{figure}
    \centering
    \includegraphics[width=1\linewidth]{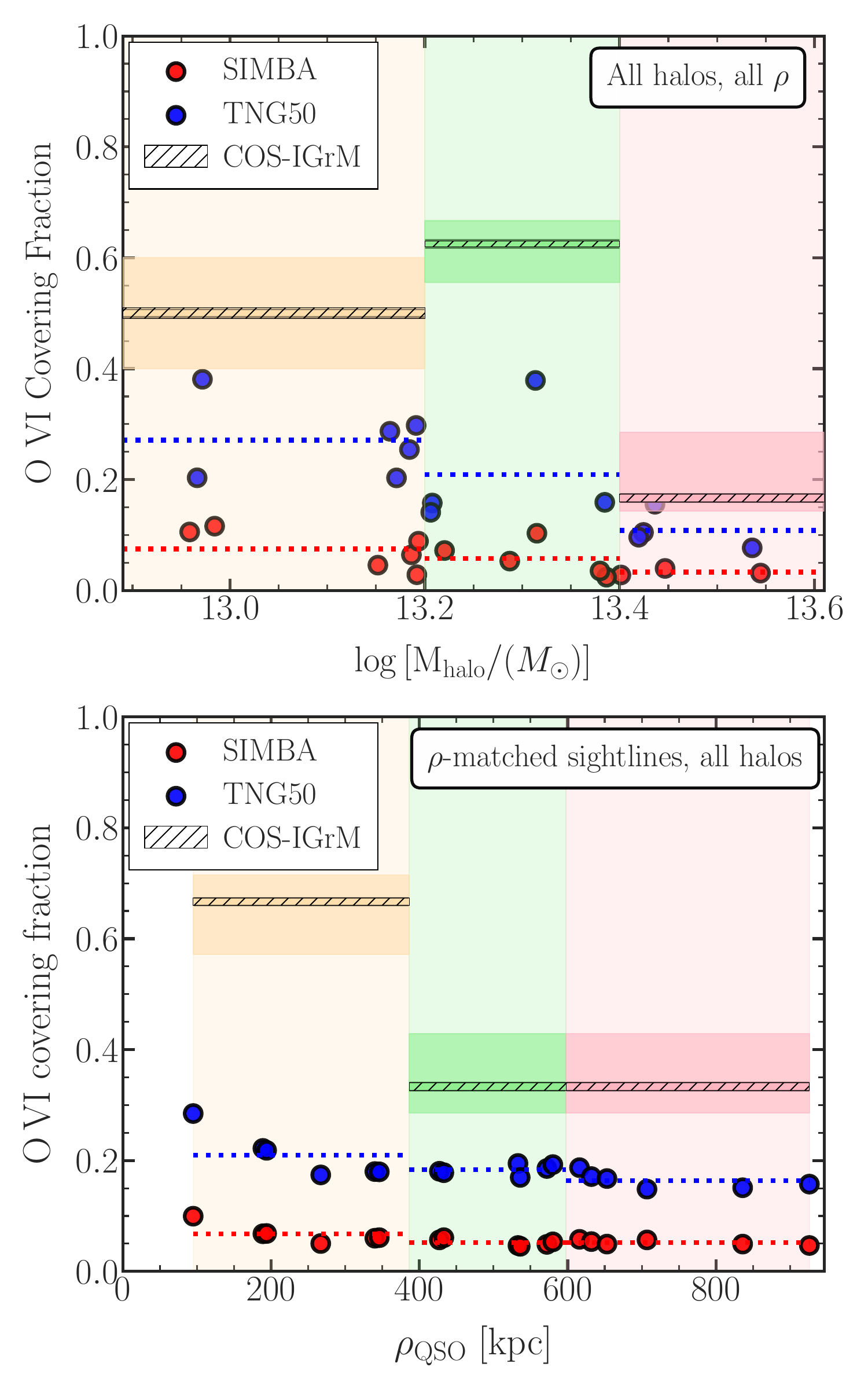}
    \caption{O\,\textsc{vi} covering fractions for the TNG50 and SIMBA groups compared to the COS-IGrM survey. \emph{Top:} Covering fraction as a function of group halo mass. Blue points represent TNG50 groups and red points represent SIMBA groups. The light shaded regions denote three halo mass bins, each containing six observational sightlines from the COS-IGrM survey. Solid hatched lines within each bin indicate the covering fractions derived from the observations, while the dotted lines show the mean covering fractions of the simulated groups in the corresponding bins. The dark shaded regions around the hatched lines represent the $1\sigma$ uncertainty on the COS-IGrM $f_{\text{OVI}}$ in each bin. \emph{Bottom:} Covering fraction as a function of impact parameter. For each COS-IGrM sightline, 2000 closest sightlines (selected uniformly from all simulated groups) are used to construct a matched sample.}
    \label{fig:covering_fraction}
\end{figure}

In the top panel of Fig.~\ref{fig:covering_fraction}, we compare the O\,\textsc{vi} covering fraction (\( f_{\text{OVI}} \)) of the TNG50 and SIMBA group samples with observations from the COS-IGrM survey, as shown in Fig.~\ref{fig:covering_fraction}. The plot is divided into three halo mass bins, each containing six observational sightlines:  
(i) \( 12.85 \leq \log M_{\text{halo}} < 13.32 \),  
(ii) \( 13.32 \leq \log M_{\text{halo}} < 13.40 \), and  
(iii) \( 13.40 \leq \log M_{\text{halo}} < 13.61 \).  
Within these bins, the observed covering fractions are represented by solid hatched lines, while the mean values from the simulations are shown as dotted lines. The COS-IGrM survey has an average \( f_{\text{OVI}} \) of \( 44 \pm 5\% \), whereas the overall mean covering fraction for TNG50 is \(20.62 \pm 2.56\%\) and for SIMBA is only \(5.98 \pm 0.82\%\). The uncertainties represent the \(1\sigma\) uncertainties, estimated by bootstrap resampling. The bottom panel of Fig.~\ref{fig:covering_fraction} presents the O\,\textsc{vi} covering fraction measured using sightlines matched in impact parameter to those of the COS-IGrM survey. For each observed sightline, the 2000 closest simulated sightlines, selected evenly from all simulated groups, were used to construct a corresponding matched sample. The resulting covering fractions remain substantially lower than observed, with $f_{\mathrm{OVI}} = 18.57 \pm 0.71\%$ for \textsc{TNG50} and $5.74 \pm 0.29\%$ for \textsc{SIMBA}. We note that the underprediction of detectable O\,\textsc{vi} persists even when the simulations are sampled to closely follow the impact parameter distribution of the observational sightlines.

\begin{figure}
    \centering
    \includegraphics[width=1\linewidth]{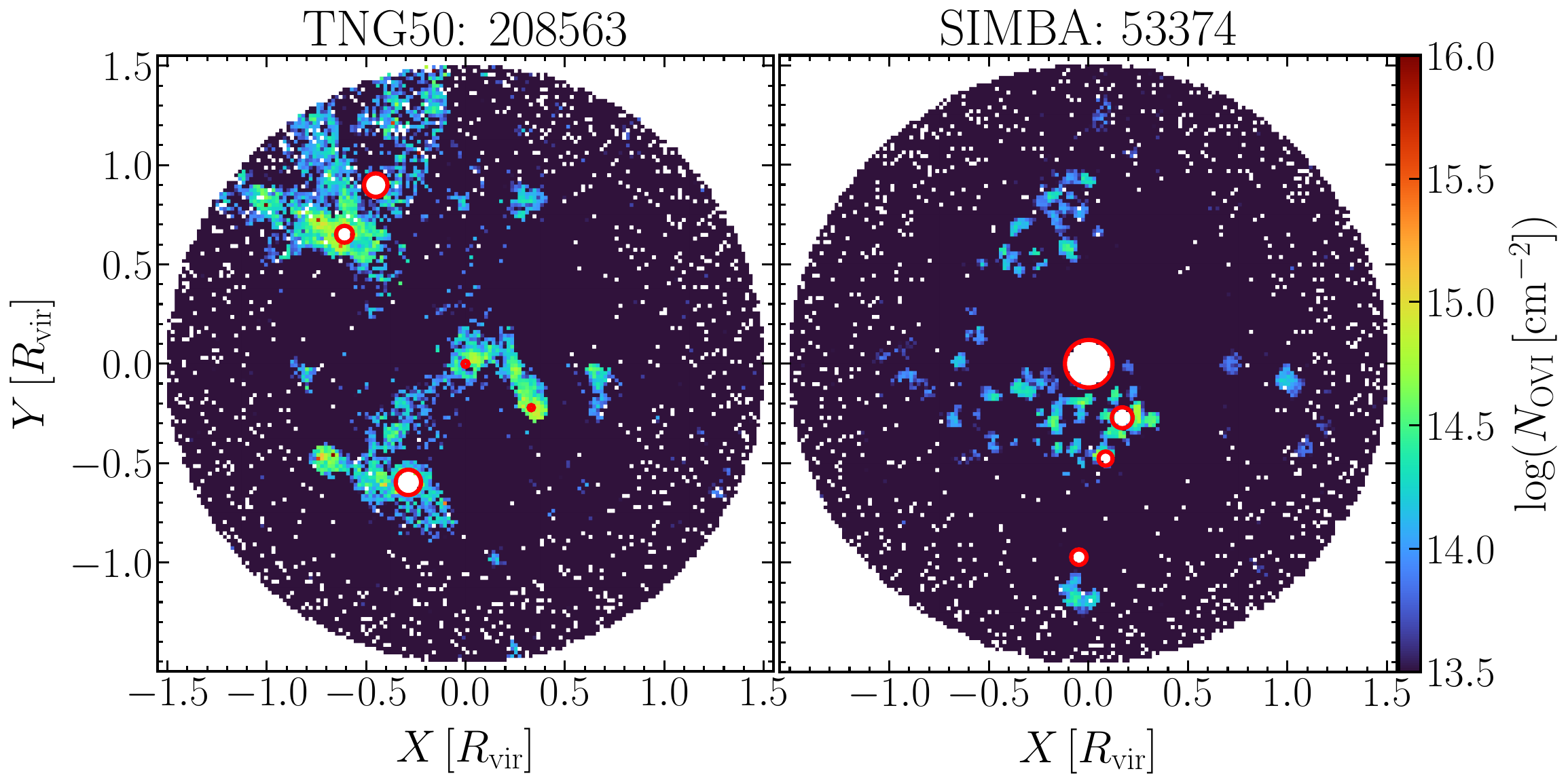}
    \caption{Spatially resolved maps of the projected  O\,\textsc{vi} column density for representative galaxy groups selected from the TNG50 (left) and SIMBA (right) simulations having similar \(M_{200c}\). Sightlines are binned to pixels and colors indicate the median \(\log N(\mathrm{O\,VI})\) within each pixel; White pixels represent regions with no sightline coverage, and red circles denote member subhalos with \(L \geq L^*\), where their radii correspond to \(3\,R_{\mathrm{1/2,\star}}\). }
    \label{fig:group_comparison}
\end{figure}

For a visual comparison of the detectable O\,\textsc{vi} distribution between TNG50 and SIMBA groups, we present binned O\,\textsc{vi} column-density maps in Fig.~\ref{fig:group_comparison} for a representative pair of TNG50 and SIMBA groups with nearly identical halo masses (\(M_{\text{halo}}\simeq 10^{13.18}\,M_{\odot}\)). Pixel colour represents the median \(\log N(\mathrm{O\,VI})\) of all sightlines passing through the pixel and white pixels contain no sightlines. The colour scale is truncated at \(\log N(\mathrm{O\,VI}) = 13.5\), corresponding to the derived lower detection limit of our study. The TNG50 group exhibits a considerably larger fraction pixels above this detection threshold compared to SIMBA, along with the presence of more higher-column-density absorbers. We confirm that a similar trend is observed across all TNG50 and SIMBA groups, although here we only illustrate a single pair for clarity.  Another key observation from the complete sample is that in all the TNG50 groups, O\,\textsc{vi} is consistently detected near the group center; the same is not true for SIMBA groups, which instead exhibit relatively diffuse distribution of detectable gas.

\begin{figure*}[htbp]    
  \centering
  \begin{subfigure}[b]{0.48\textwidth}
      \includegraphics[width=\linewidth]
        {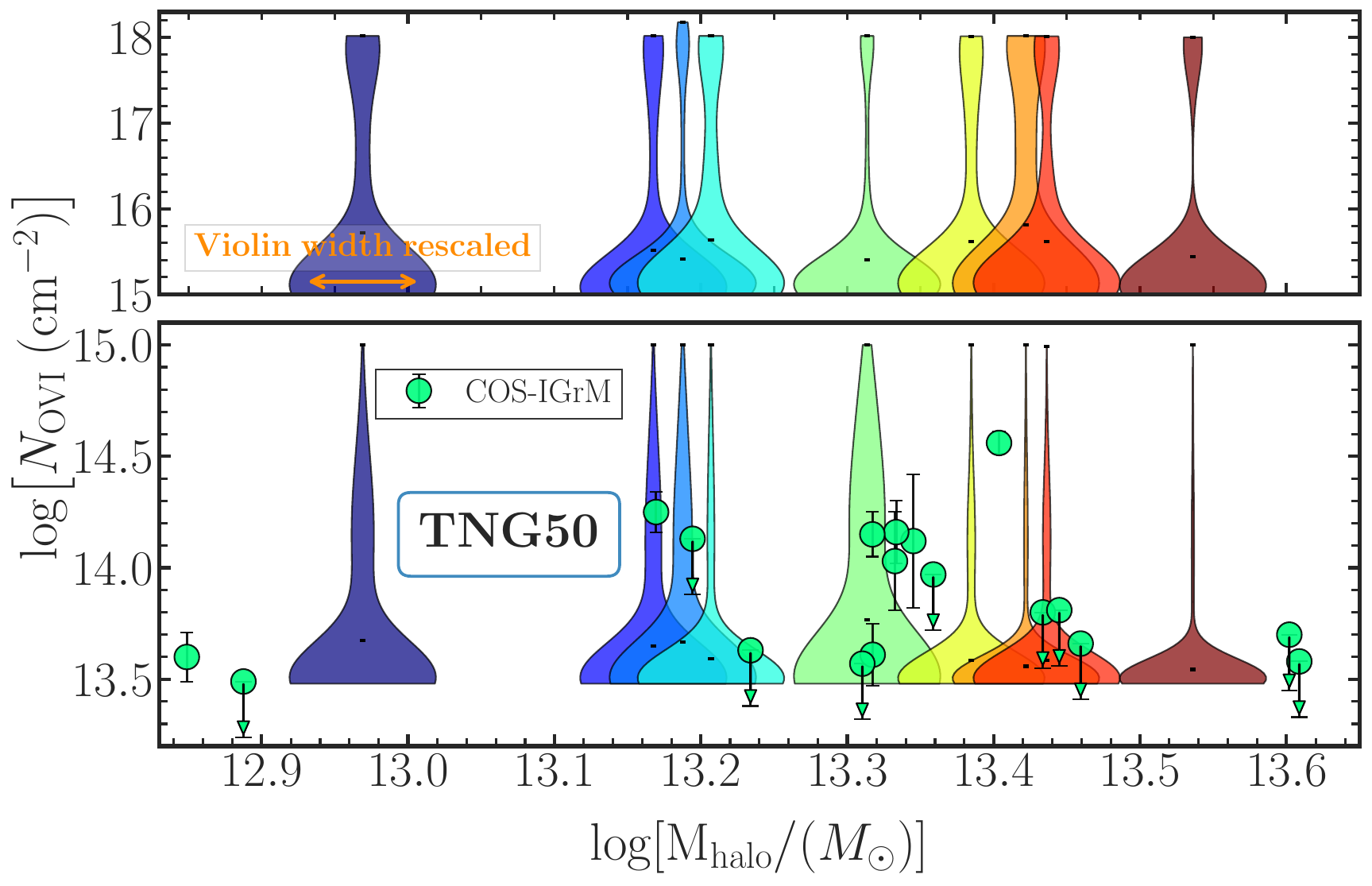}
      \phantomcaption       
      \label{fig:N_vs_M_halo}
  \end{subfigure}%
  \begin{subfigure}[b]{0.48\textwidth}
      \includegraphics[width=\linewidth]{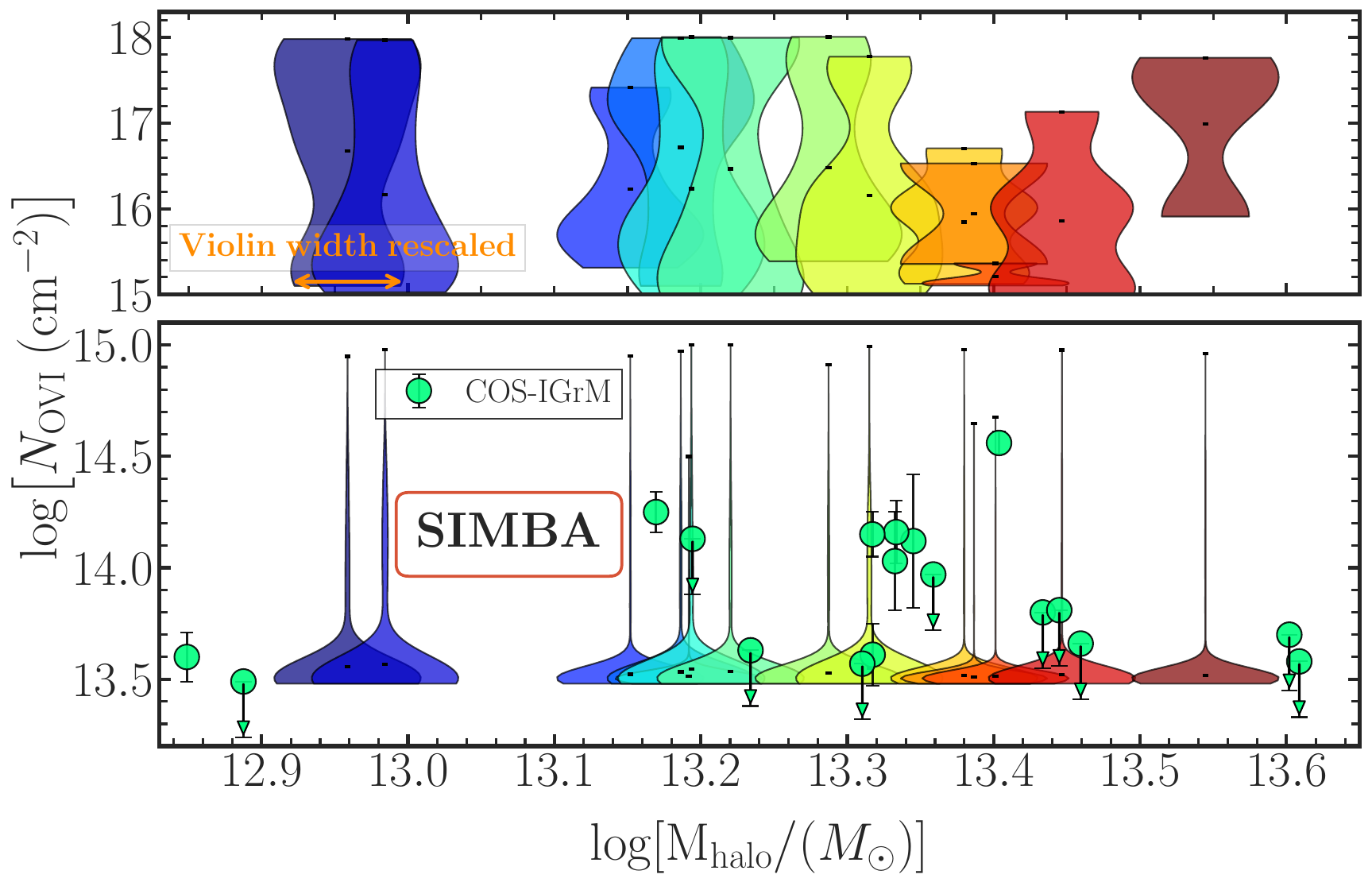}
      \phantomcaption
      \label{fig:N_vs_M_halo_SIMBA}
  \end{subfigure}

  \vspace{-1em}

  \caption{ (Left) \textbf{TNG groups:} (Bottom) O\,\textsc{vi} column-density
    distribution as a function of group halo mass, compared to observations
    from the COS-IGrM survey. Fourteen groups are binned into nine bins for
    clear visualization with assignments: Bin 1 (294515, 313950), Bin 2
    (305020, 208563), Bin 3 (288932, 277688), Bin 4 (264620, 283512),
    Bin 5 (247945), Bin 6 (239843), Bin 7 (231369, 199226), Bin 8 (219842),
    Bin 9 (188893).  The halo mass shown is the mean halo mass of each bin.
    Filled green circles denote COS-IGrM data \citep{mccabe_2021}. (Top) A
    zoomed-in region with $\log[\mathrm{N(O\,\textsc{vi})}]>15$ visualizes
    the distribution of strong and saturated absorbers. (Right) Same as left but for SIMBA groups without any binning.%
  }
  \label{fig:Combined_NOVI}
\end{figure*}

For the TNG50 and SIMBA sample, we examine the distribution of O\,\textsc{vi} absorber column densities as a function of halo mass, as shown in Fig.~\ref{fig:N_vs_M_halo} (left and right panel respectively). Each violin plot represents the smoothed density distribution of the column densities for detected absorbers and upper limits within each halo mass bin. The mean column density for each bin is shown with a small black marker inside the violin. A rescaled region highlighting high column density absorbers with \( \log\mathrm{N} > 15 \) are shown in the upper panel. A \textit{Spearman} correlation test for TNG, shows a correlation coefficient of \(\rho = -0.14\) with a $p$-value of $0.0$. While the result is statistically significant as indicated by small p value, the anti-correlation is weak (small \(\rho\)) , suggesting that halo mass has little effect on the distribution of column densities. A two-sample Kolmogorov--Smirnov (KS) test comparing the cumulative distributions of the O\,\textsc{vi} column densities from the TNG groups and the COS-IGrM observations yielded a KS statistic of 0.769 with a $p$-value $\ll 10^{-3}$. The large KS statistic, together with the very low $p$-value, indicates that the two distributions are statistically different. We draw a similar conclusion based on the test done on the SIMBA sample.

\begin{figure*}[htbp]    
  \centering
 
  \begin{subfigure}[b]{0.48\textwidth}
      \includegraphics[width=\linewidth]
        {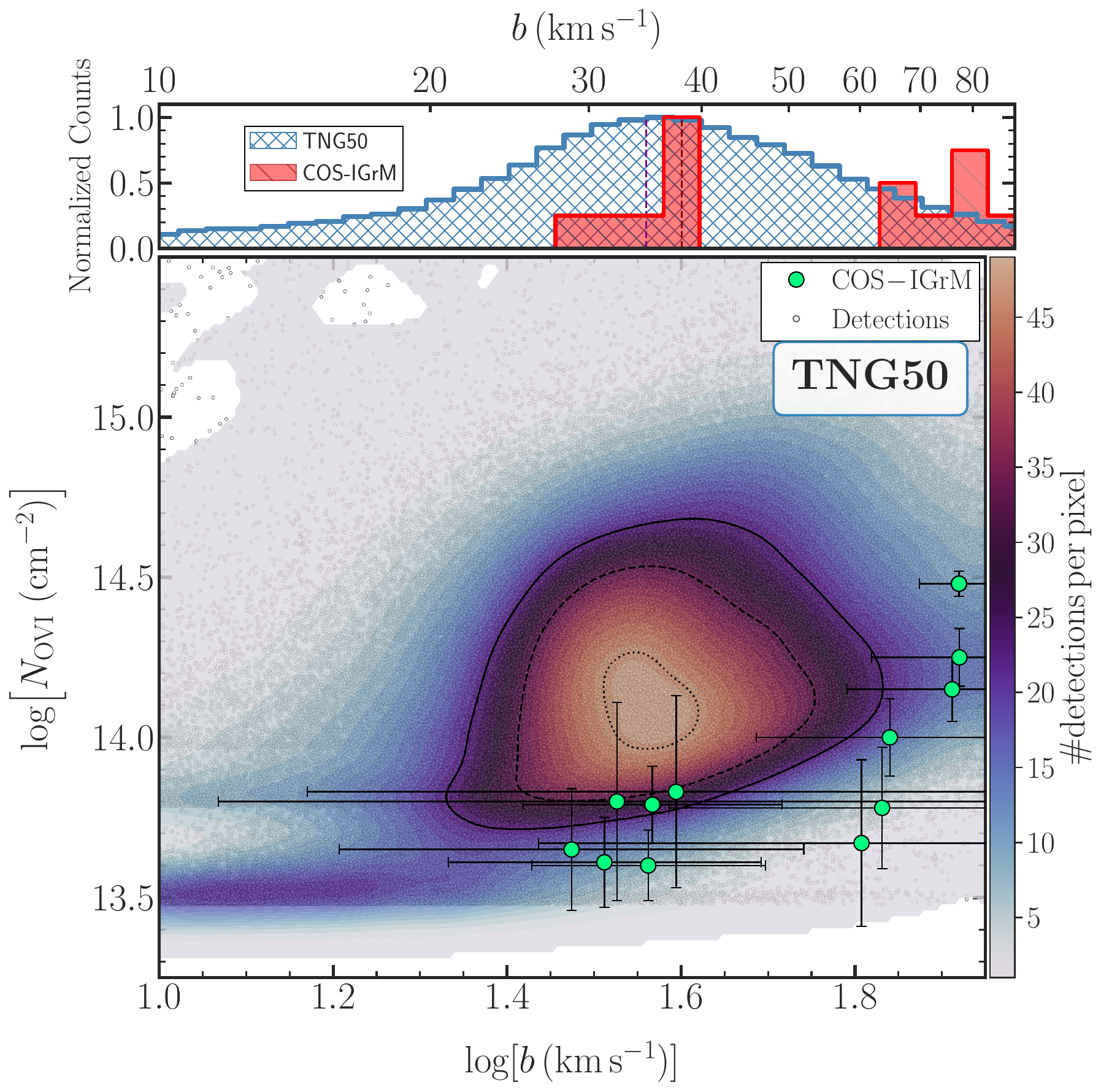}
      \phantomcaption
      \label{fig:logN_vs_logb}
  \end{subfigure}%
  \begin{subfigure}[b]{0.48\textwidth}
      \includegraphics[width=\linewidth]
        {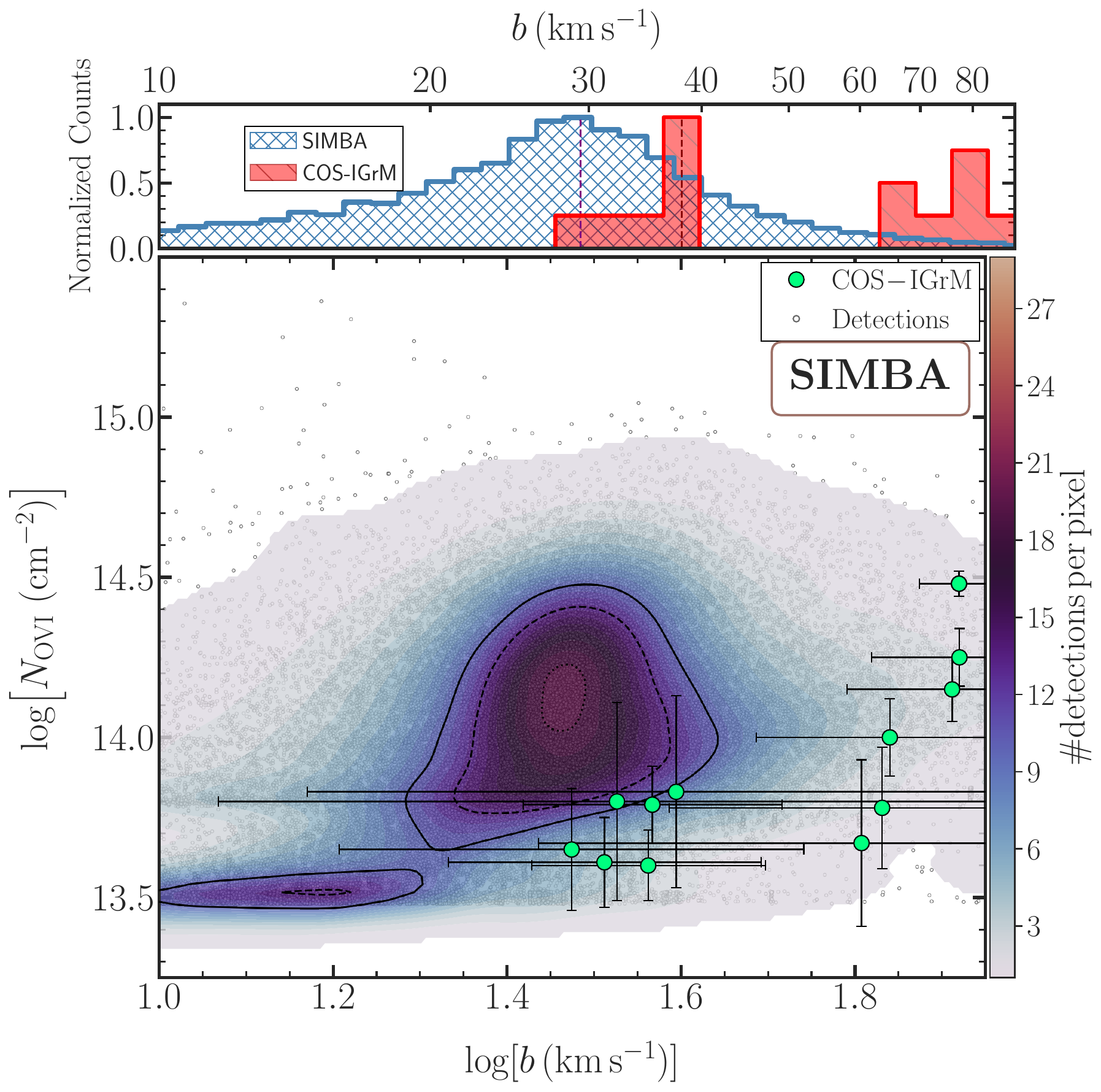}
      \phantomcaption
      \label{fig:NOVI_doppler_SIMBA}
  \end{subfigure}
  \vspace{-1em}
  \caption{
    (Left) O\,\textsc{vi} column density of detected absorbers in synthetic
    sightlines from TNG50 groups versus Doppler (\(b\)) parameter.  The
    upper panel shows the normalized histogram of \(b\) for the same
    detections. (Right) Same as left but for SIMBA groups.}
  \label{fig:Combined_logN_logb}
\end{figure*}

The distribution of column density for the detected absorbers as as a function of their Doppler $b$ parameter is shown in Figure~\ref{fig:Combined_logN_logb}. The upper sub-panel displays the distribution of the $b$ parameter values for all detected components in blue. The histogram was normalized to have a peak value of 1. Two artifact peaks result from the predefined upper and lower limits set during the Voigt profile fitting pipeline at $6\,\mathrm{kms^{-1}}$ and $100\,\mathrm{kms^{-1}}$. Excluding these boundary bins, the intrinsic peak of the TNG50 distribution is found at approximately $35 \,\mathrm{kms^{-1}}$ while for SIMBA, it is $29 \,\mathrm{kms^{-1}}$. For comparison, normalized histogram of $b$ parameter for COS-IGrM observations is presented in red.

The $\log N$ versus $\log b$ plot is an important diagnostic for understanding the physical and kinematic state of gas. The broader widths observed for O\,\textsc{vi} in both the simulations  groups suggest either higher gas temperatures or significant non-thermal broadening processes.

At the peak collisional ionization equilibrium (CIE) temperature for O\,\textsc{vi} ($T \approx 2.5\times10^5\,\mathrm{K}$; \citealt{Gnat_2007,Oppenheimer_2013}), the purely thermal Doppler width is roughly $16\,\mathrm{km\,s^{-1}}$, significantly lower than the observed width. Thus, the much larger observed width (TNG $\sim35\,\mathrm{km\,s^{-1}}$ ; SIMBA $\sim29\,\mathrm{km\,s^{-1}}$) implies a dominant non-thermal broadening component. This non-thermal component may arise from unresolved velocity components, and/or may indicate inflows, outflows or turbulent motions.

\begin{figure*}[htbp]
    \centering

    \begin{subfigure}[b]{0.49\textwidth}
        \centering
        \captionsetup{labelformat=empty}
        \includegraphics[width=\linewidth]{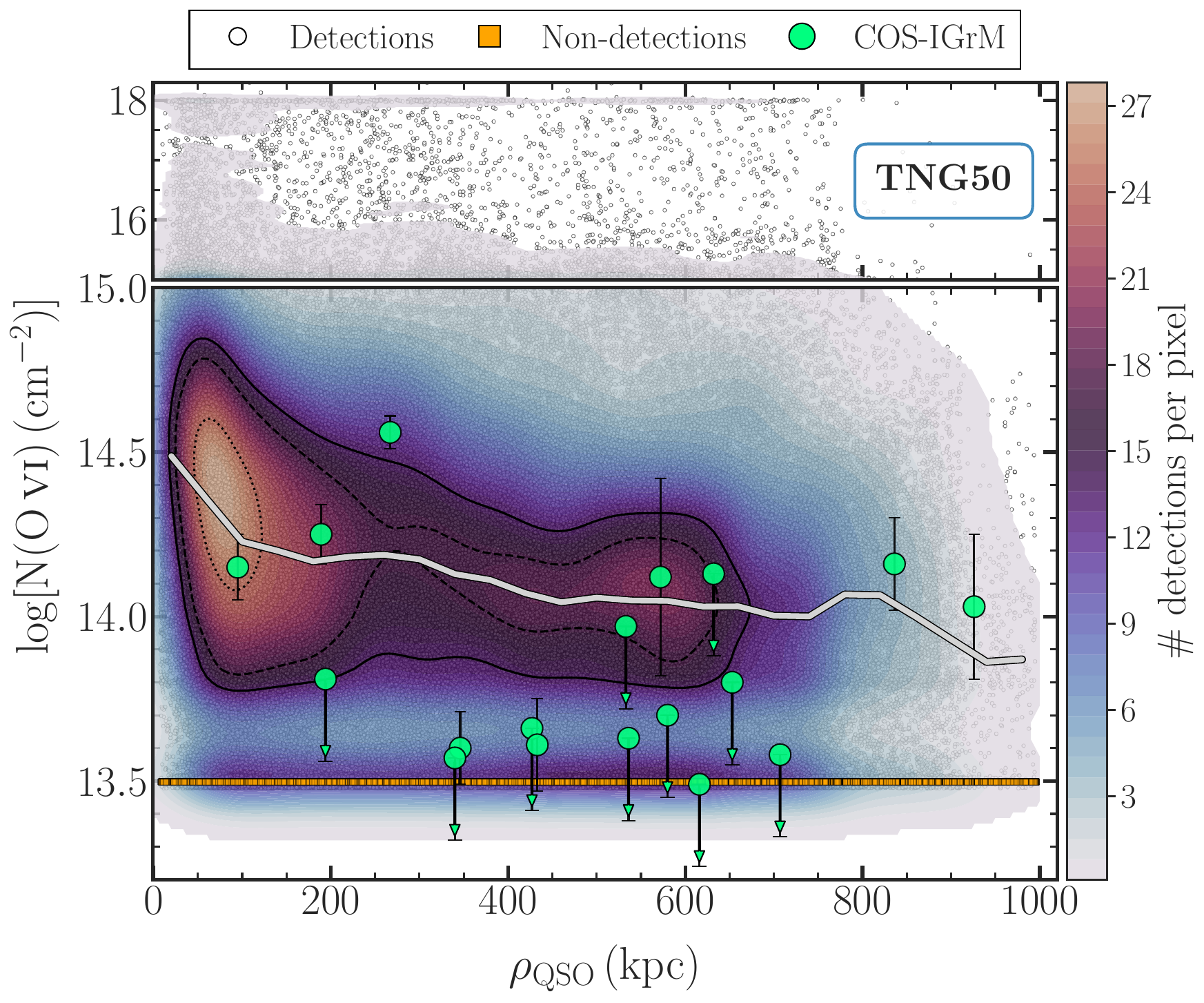}
    \end{subfigure}
    \hfill
    \begin{subfigure}[b]{0.49\textwidth}
        \centering
        \captionsetup{labelformat=empty}
        \includegraphics[width=\linewidth]{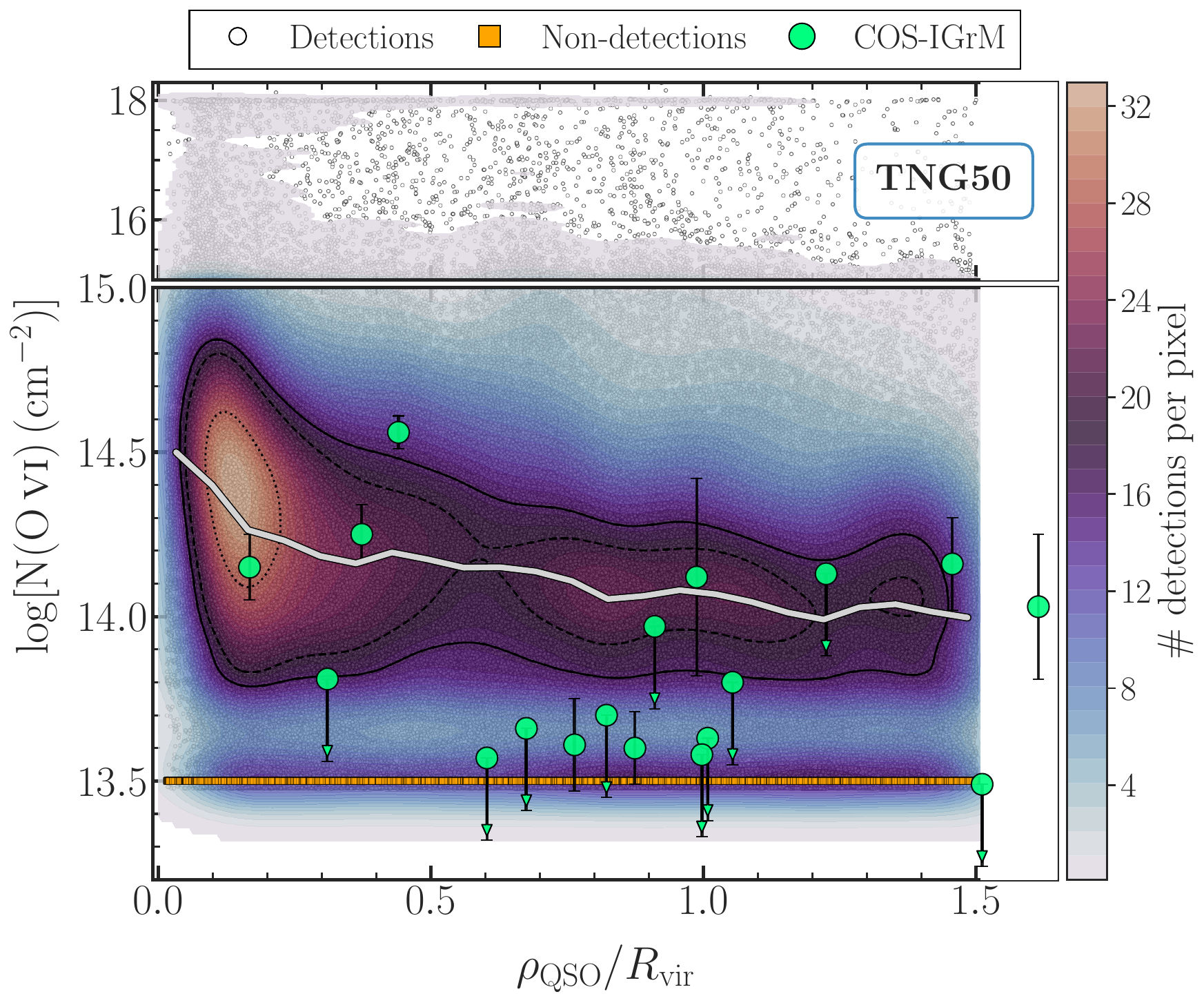}
    \end{subfigure}

    \begin{subfigure}[b]{0.49\textwidth}
        \centering
        \captionsetup{labelformat=empty}
        \includegraphics[width=\linewidth]{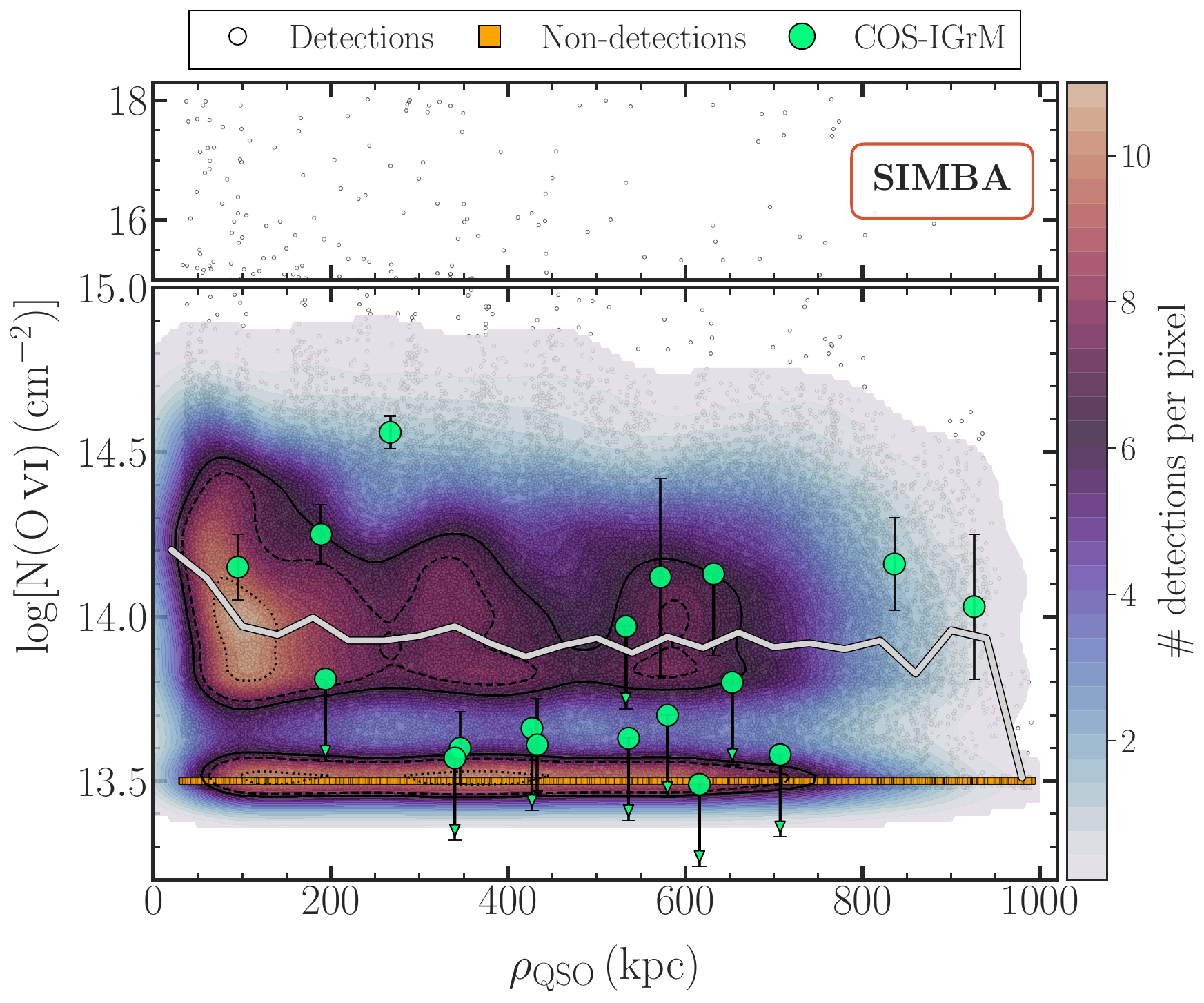}
    \end{subfigure}
    \hfill
    \begin{subfigure}[b]{0.49\textwidth}
        \centering
        \captionsetup{labelformat=empty}
        \includegraphics[width=\linewidth]{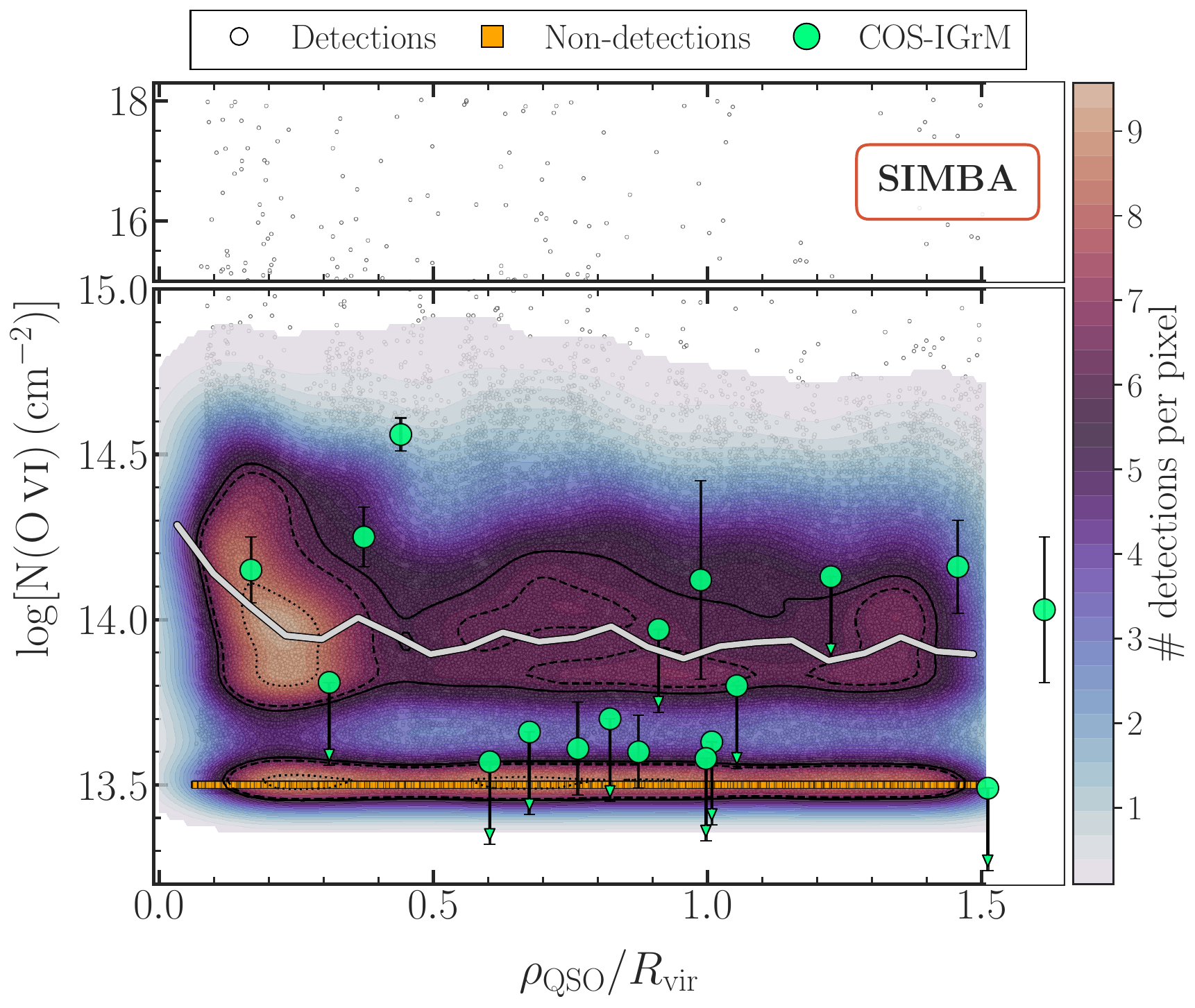}
    \end{subfigure}
    \vspace{-0em}
    
    \caption{\textit{Top row:} O\,\textsc{vi} column density for all synthetic sightlines
    in the selected TNG50 groups as a function of (left) impact parameter from the
    group centre and (right) the same quantity normalised by the virial radius
    (\(R_{200c}\)). Grey circles denote detections, orange squares show
    \(3\sigma\) upper limits for non-detections, green points correspond to
    COS-IGrM observations, and the solid grey line shows the running median of the
    column densities for detected absorbers. Solid, dashed, and dotted contours
    represent the 50\%, 68\%, and 95\% confidence levels of the 2D KDE for
    detections. Smaller sub-panel at the top shows strong \(\log(N_{\text{OVI}}) > 15\) absorbers.
    \textit{Bottom row:} The same diagnostics for \textsc{SIMBA} groups.}
    \label{fig:Combined_ColumnDensity_vs_rho_TNG50_SIMBA}
\end{figure*}

Figure \ref{fig:Combined_ColumnDensity_vs_rho_TNG50_SIMBA} illustrates the O\,\textsc{vi} column density as a function of the impact parameter of the sightline in $kpc$ as well as normalized by the virial radius, respectively. The contours show two-dimensional kernel density which represent the concentration of the absorbers. For TNG groups, we observe that most of the absorbers are located near the 
center of the group at \(\rho \lesssim 200~\mathrm{kpc}\). 
We see a flat trend in the distribution of absorbers for 
\(0.6R_{\mathrm{vir}} \lesssim \rho \lesssim 1.5R_{\mathrm{vir}}\) in 
Fig.~\ref{fig:Combined_ColumnDensity_vs_rho_TNG50_SIMBA}. 
At smaller impact parameters \((0.1R_{\mathrm{vir}} \lesssim \rho \lesssim 0.6R_{\mathrm{vir}})\), 
there is a modest increase in absorbers with higher column densities at 
lower \(\rho\). Saturated absorbers with \(\log N > 15\) are concentrated near the group center, as seen for \(\rho \lesssim 100~\mathrm{kpc}\) or equivalently \(\rho \lesssim 0.3R_{\mathrm{vir}}\). This is in alignment with the observations presented in \cite{mccabe_2021}. SIMBA groups exhibit a similar radial trend. However, the fraction of absorbers with 
\(\log N > 15\) is significantly smaller compared to TNG50. For the SIMBA groups, the overall trend is flatter as compared to TNG and there is only a nominal increase in overall increase in absorbers at $\rho<0.4 R_{vir}$.

\begin{figure}[htbp]
\centering

    \includegraphics[width=\linewidth, trim={0cm 1.5cm 0cm 1cm}, clip]{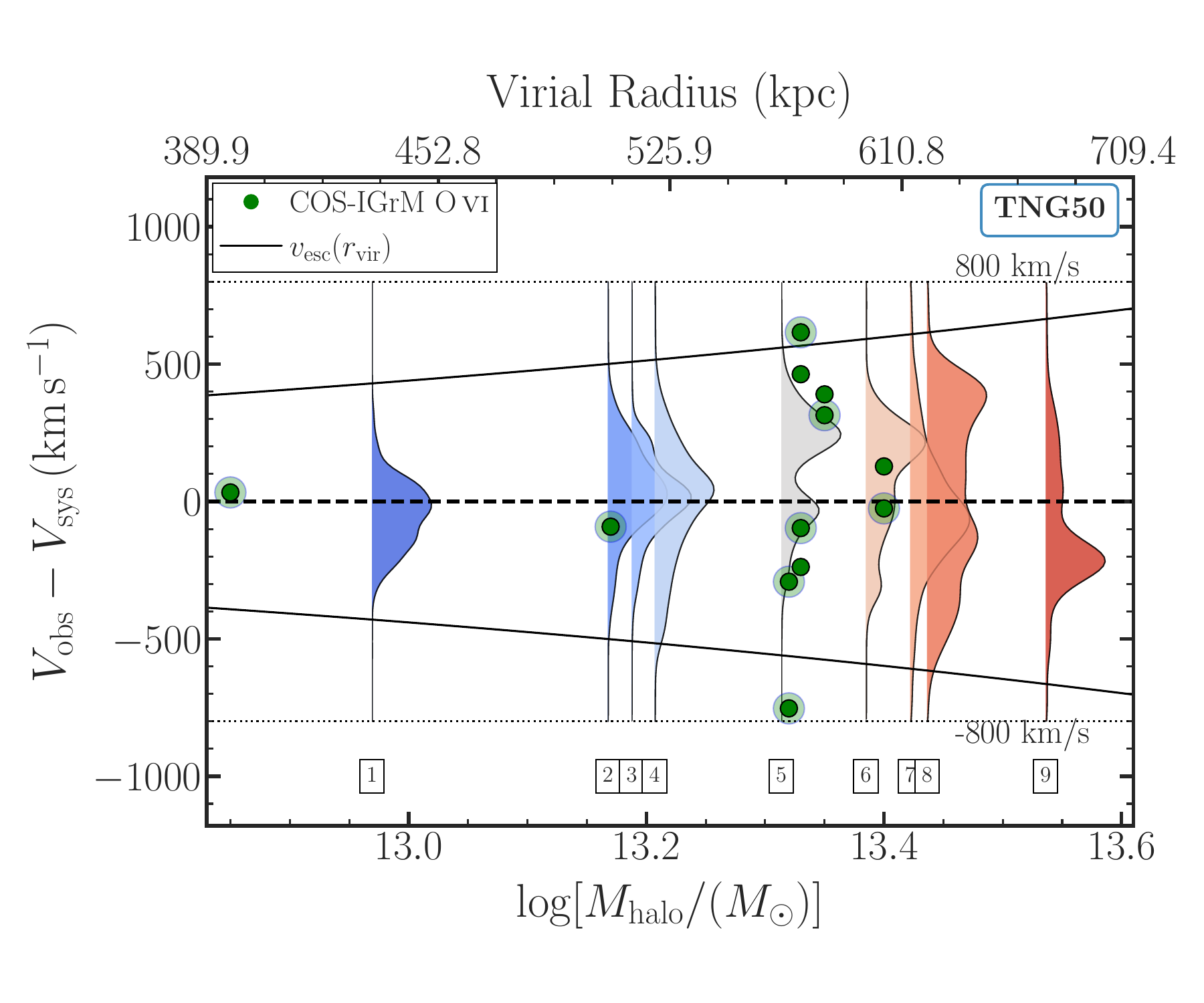}

    \includegraphics[width=\linewidth, trim={0cm 1cm 0cm 1.5cm}, clip]{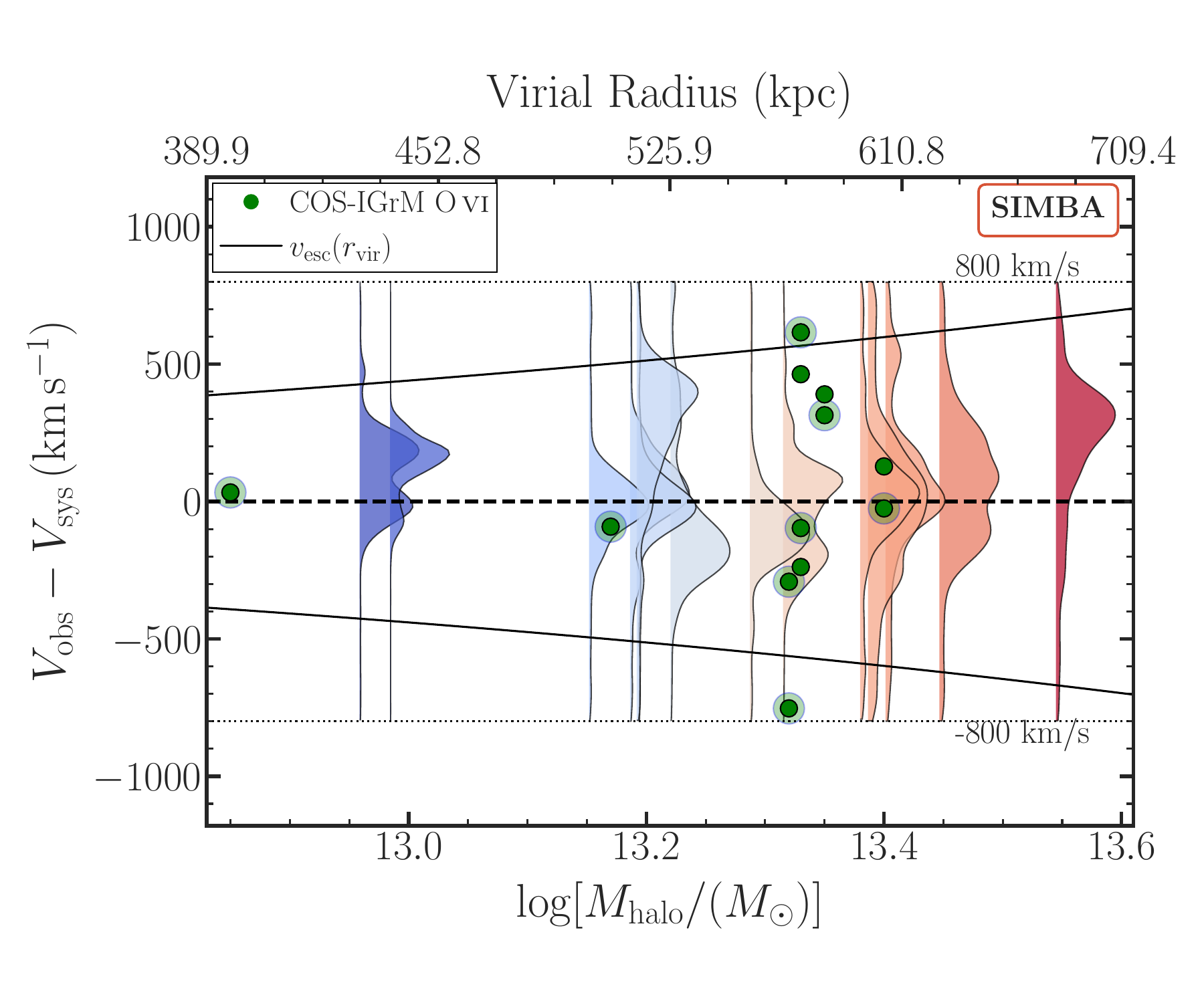}
    \caption{Line-of-sight velocity relative to the group centre for all synthetic
    sight-lines containing detected O\,\textsc{vi} absorbers, shown as a function of
    group halo mass. \textit{Top:} TNG50 groups, where fourteen groups are
    binned into nine bins for clear visualisation using the same binning scheme
    as Figure~\ref{fig:N_vs_M_halo}. Violin plots represent the density
    distribution of detected velocities, the solid black line marks the escape
    velocity of each group’s gravitational potential, and green points denote
    COS-IGrM absorbers (with the highest column densities highlighted by lighter
    circles). \textit{Bottom:} Same diagnostic for \textsc{SIMBA} groups.}
    \label{fig:vel_vs_halo_mass_combined}
\end{figure}

Figure~\ref{fig:vel_vs_halo_mass_combined} shows the velocity distribution of detected absorbers in TNG50 (top) and SIMBA (bottom)  groups relative to their systemic velocities. 
The distribution is represented using half-violin plots, with the binning scheme following the same approach as in Figure~\ref{fig:N_vs_M_halo}\footnote{The width of the violin represents only the velocity distribution and does not indicate any dispersion in halo mass.}. 
The dashed black line at \(V_{\mathrm{obs}} - V_{\mathrm{sys}} = 0\) represents the systemic velocity of the group, while the solid black lines indicate the escape velocity as a function of the virial radius and halo mass. 
Green points represent absorbers from the COS-IGrM survey, with the highest column density absorbers within a given sightline highlighted by a larger halo around the point. The virial radius (\(R_{\mathrm{vir}}\)) and the escape velocity (\(v_{\mathrm{esc}}\)) 
were estimated following the prescription provided in  \citet{mccabe_2021}.

In the TNG50 sample, we find that \(98.6\%\) of the absorbers are gravitationally bound to their group halos, with a dispersion of \(0.7\%\) across individual groups (where the dispersion reflects the standard deviation of the bound fraction among groups) as compared to \(83\%\) (10 out of 12) of such absorbers in the COS-IGrM survey. In SIMBA sample we observe the bound fraction to be \(94.2\%\) with a dispersion of \(2.5\%\) 
We find that the absorber kinematics are generally not symmetric with respect  to the group systemic velocity. In some groups, the velocity distributions exhibit two distinct peaks, with offsets as large as 
\(\sim 250~\mathrm{km~s^{-1}}\), suggesting bulk motions within the gas.

\begin{figure}[htbp]
    \centering

    \includegraphics[width=\linewidth]{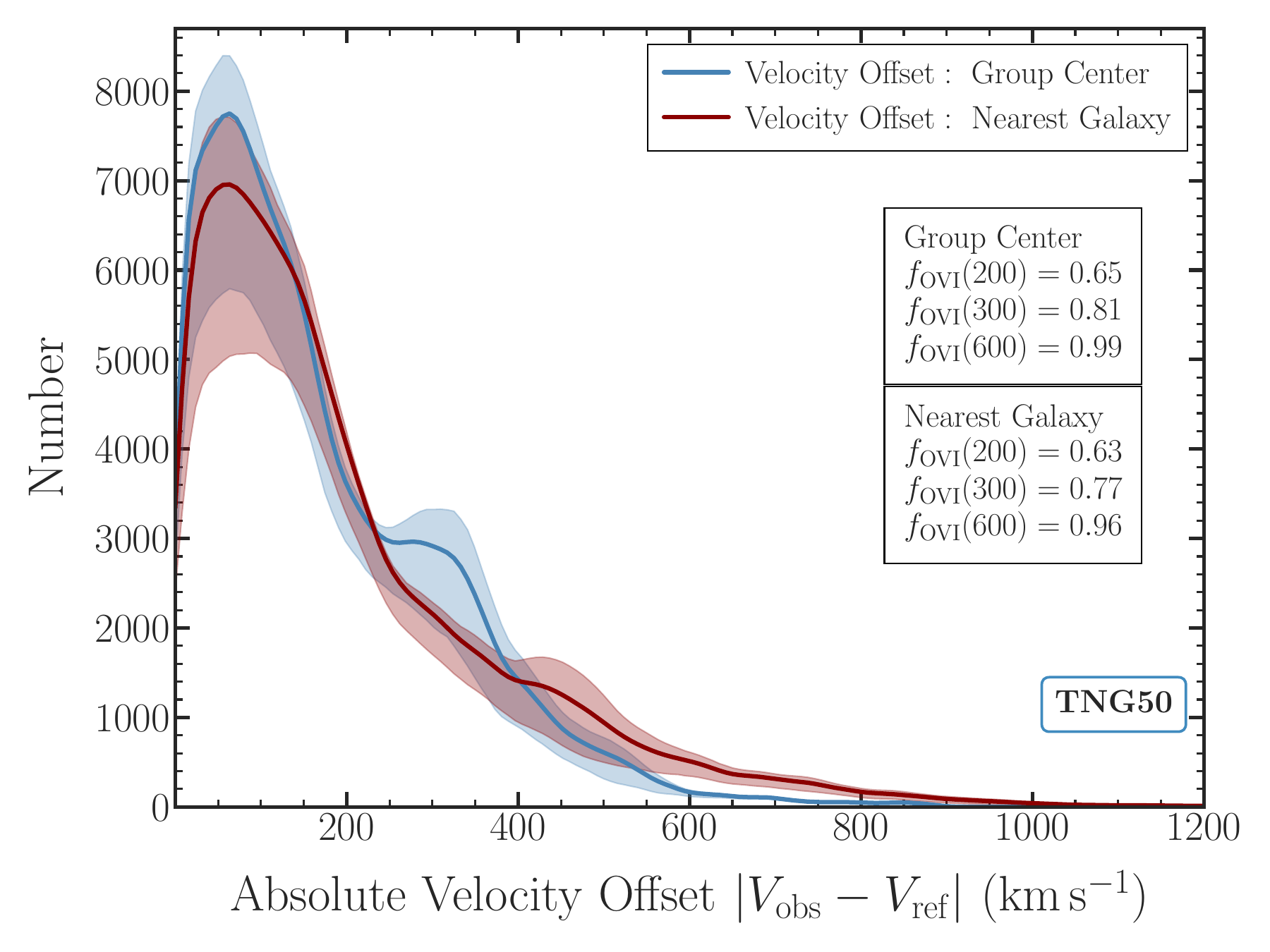}

    \vspace{0em}

    \includegraphics[width=\linewidth]{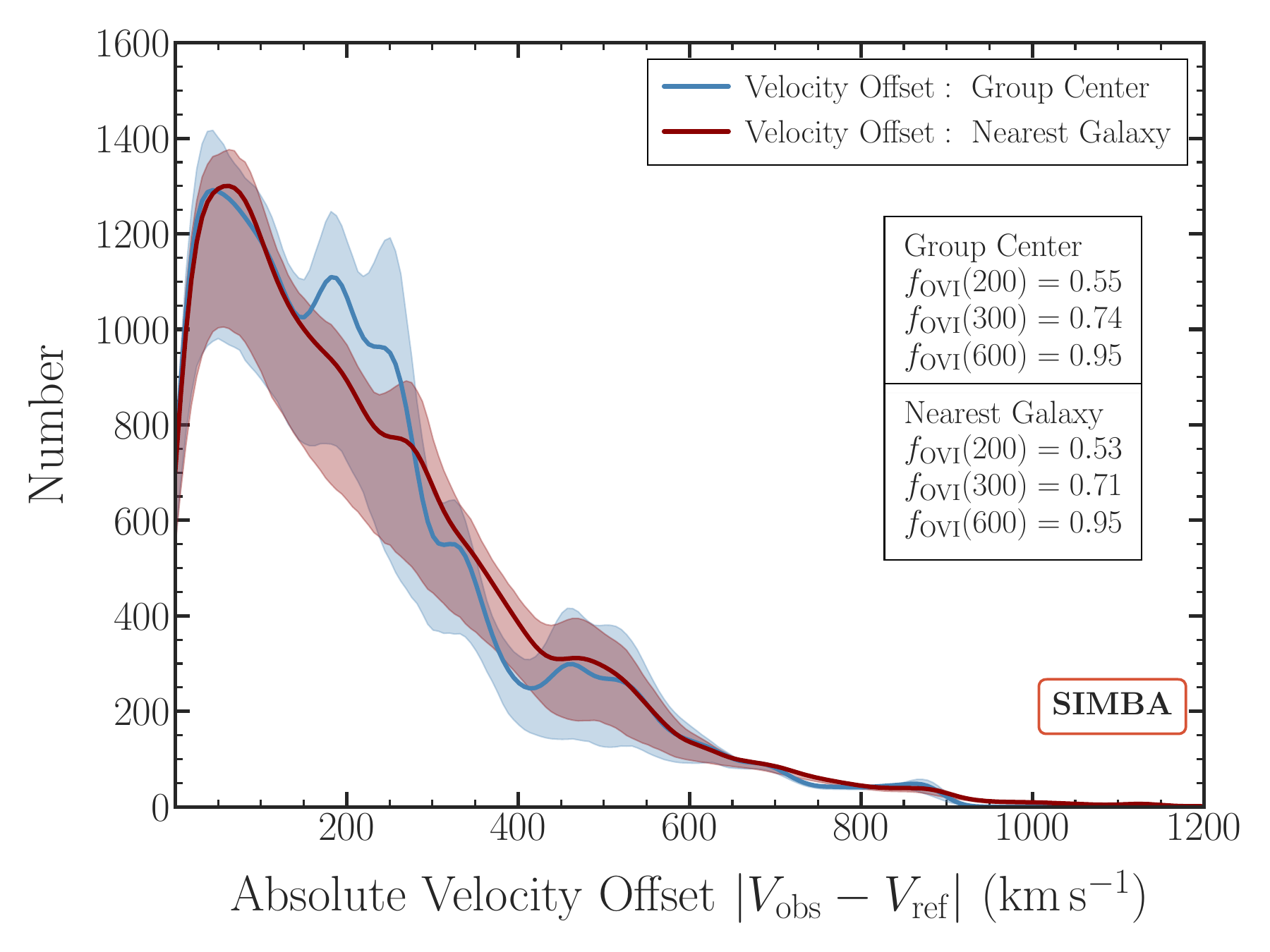}

    \caption{Absolute velocity offset of O\,\textsc{vi} detections relative to (i) the centre of the host group and (ii) the nearest galaxy to each absorber. Insets give the fractions of absorbers within 200, 300, and \(600~\mathrm{km\,s^{-1}}\) of the group center and nearest galaxy
    \textit{Top:} TNG50 groups.
    \textit{Bottom:} Same diagnostic for \textsc{SIMBA} groups.}
    \label{fig:absolute_v_distrib_combined}
\end{figure}

Figure~\ref{fig:absolute_v_distrib_combined} presents a histogram of velocity offsets for all O\,\textsc{vi} absorbers relative to both the group center and the nearest galaxy. The shaded regions correspond to the 1$\sigma$ statistical uncertainty derived from a bootstrap analysis, where each iteration randomly selects 13 out of the 14 groups (with replacement). The vast majority of absorbers are gravitationally bound to their group in both the simulations. We note that, compared to SIMBA, a larger fraction of absorbers in TNG groups lie closer to both the group systemic velocity and the velocity of the nearest galaxy, as indicated by the higher proportion within 200–600~km~s\(^{-1}\) in Fig.~\ref{fig:absolute_v_distrib_combined}. Within each simulation, however, no significant difference is seen between the distributions relative to the group and to the nearest galaxy.

\begin{figure*}[htbp]    
  \centering

  \begin{subfigure}[b]{0.48\textwidth}
      \includegraphics[width=\linewidth]
        {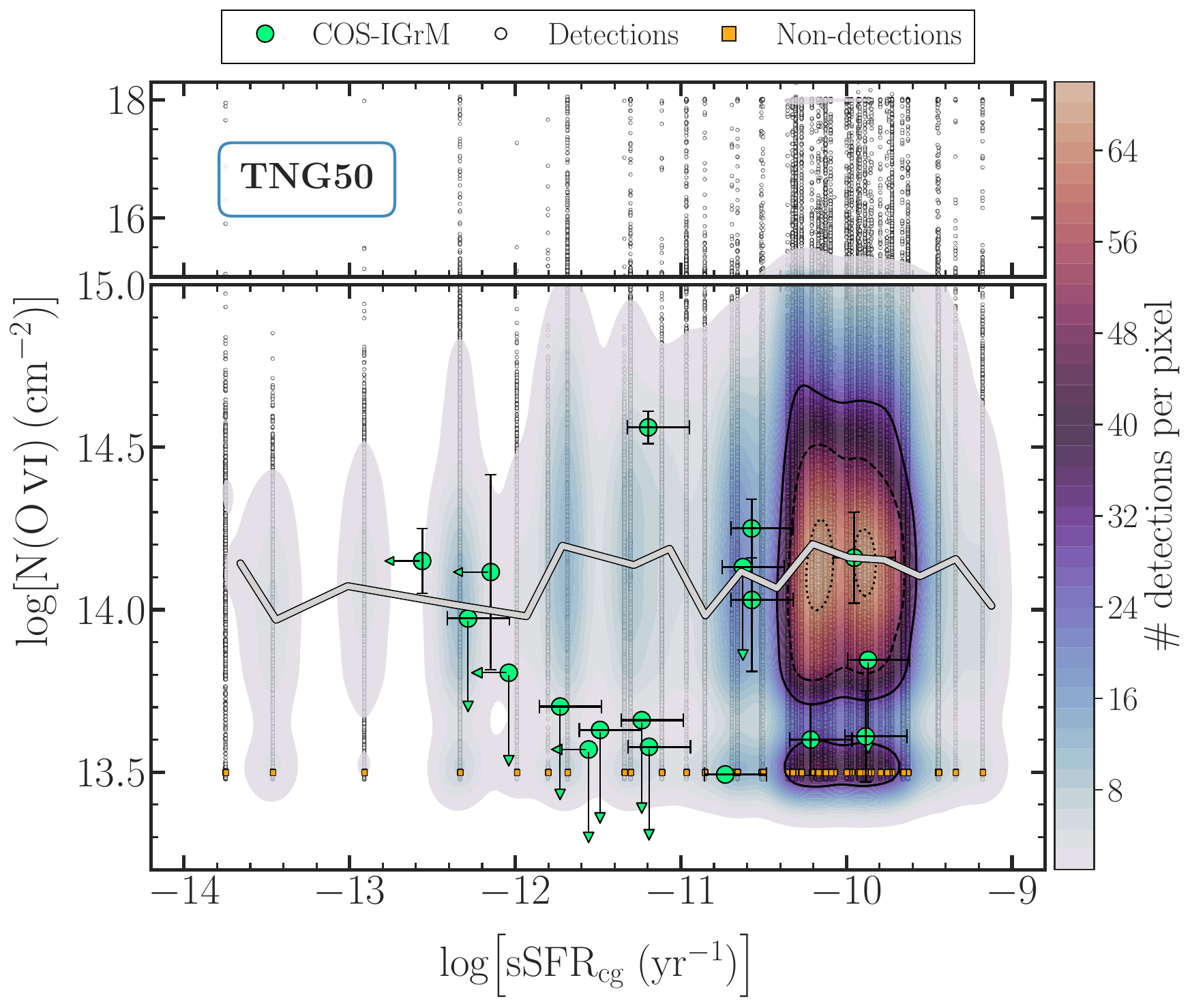}
      \phantomcaption          
      \label{fig:logN_vs_sSFR}
  \end{subfigure}%
  \begin{subfigure}[b]{0.48\textwidth}
      \includegraphics[width=\linewidth]
        {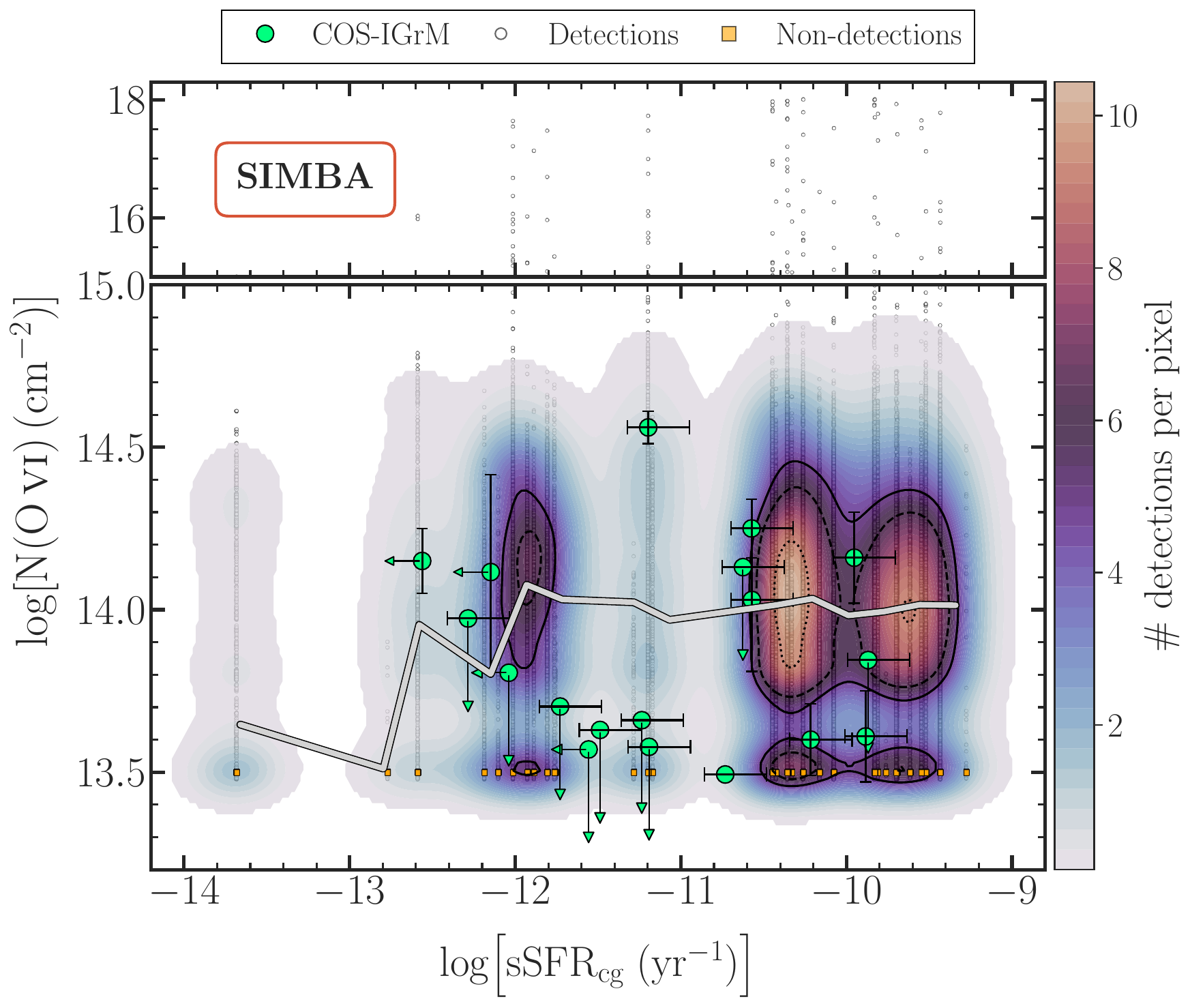}
      \phantomcaption
      \label{fig:NOVI_vs_sSFR_SIMBA}
  \end{subfigure}

  \caption{
    (Left) O\,\textsc{vi} column density of all synthetic sightlines in
    TNG50 groups versus the specific star‐formation rate of the nearest
    group member galaxy with \(L \ge L_{*}\). Scheme for the 2D KDE plot and scatter points is same as Fig. \ref{fig:Combined_ColumnDensity_vs_rho_TNG50_SIMBA}.(Right) Same as left for SIMBA groups.}
  \label{fig:Combined_logN_sSFR}
\end{figure*}

Previous O\,\textsc{vi} studies, such as the COS-Halos survey \citep{Tumlinson_2011}, have found a strong correlation between O\,\textsc{vi} absorption in the CGM and the star formation rate of galaxies. We investigate whether a similar trend is present in our sample. Figure~\ref{fig:Combined_logN_sSFR} shows the column density distribution of TNG50 (left) and SIMBA (right) absorbers as a function of the specific star formation rate (sSFR) of the nearest member galaxy with \(L \geq L_{*}\). For the TNG sample, the majority of absorbers cluster around \(\log(sSFR) \sim -10.25\) and \(\log(N_{\text{OVI}}) \sim 14\). High column density absorbers are also concentrated in the same region. Overall, we find that 83\% of all absorbers are associated with galaxies with \(\log(sSFR) > -11\), and this fraction rises to 90\% when considering only the high column density detections (\(\log(N_{\text{OVI}}) > 15\)). In the SIMBA sample, absorbers cluster around a column density of \(\log(N_{\text{OVI}}) \sim 13.8\), but are spread across three distinct sSFR: \(\log(sSFR) \sim -9.8, -10.5,\) and \(-12.2\). The much more fragmented distribution for SIMBA when compared to TNG is primarily an artifact because of the intrinsically lower number of O\,\textsc{vi} detections in SIMBA and the greater diversity in star formation properties of its member galaxies. Among all SIMBA absorbers, 57\% are associated with galaxies having \(\log(sSFR) > -11\), which decreases to 50\% when considering only high column density absorbers. 

\begin{figure*}[t]
    \centering

    \begin{minipage}[t]{0.65\textwidth}
        \centering
        \includegraphics[width=\linewidth]{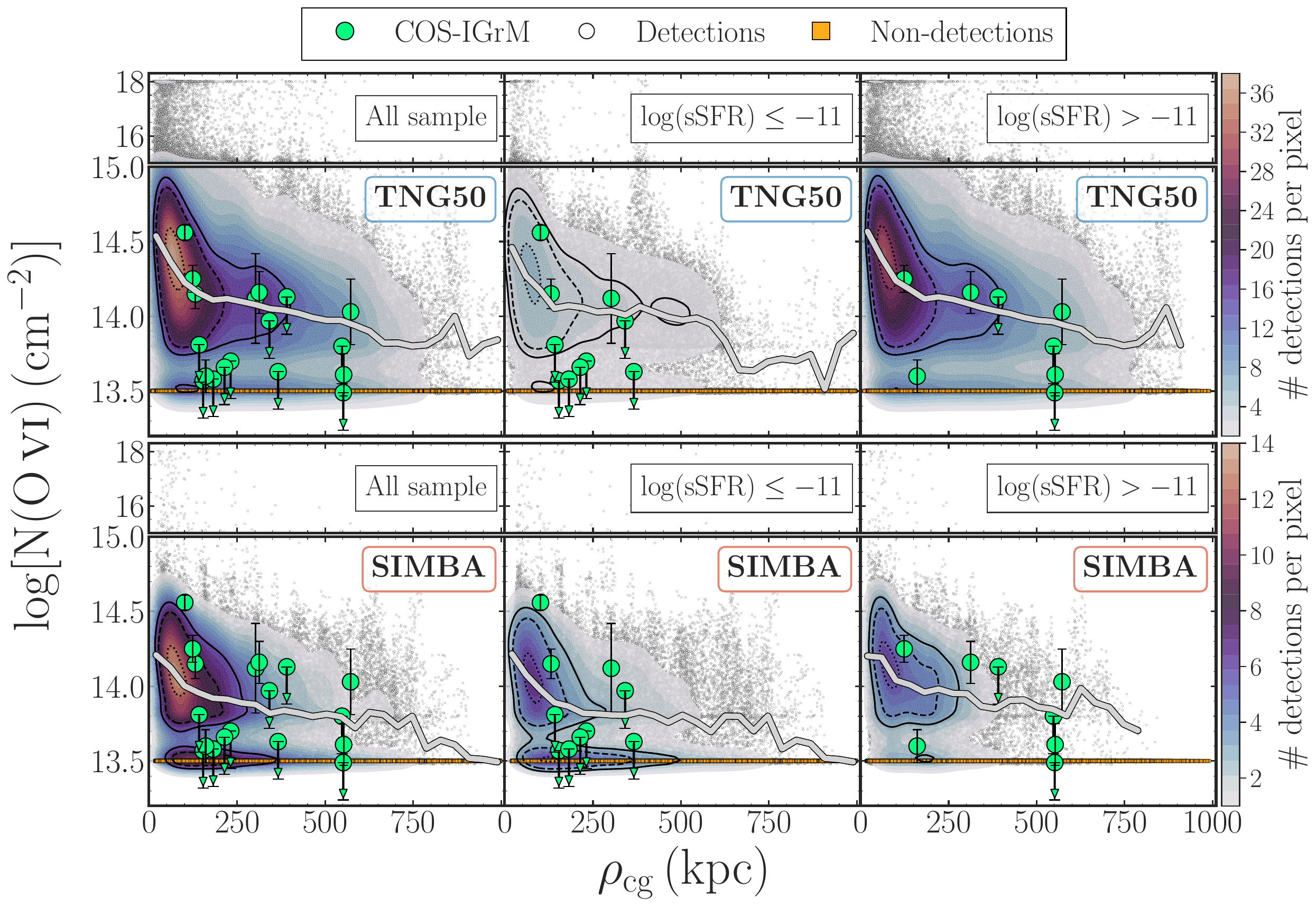}
    \end{minipage}%
    \begin{minipage}[t]{0.35\textwidth}
        \centering
        \includegraphics[width=\linewidth]{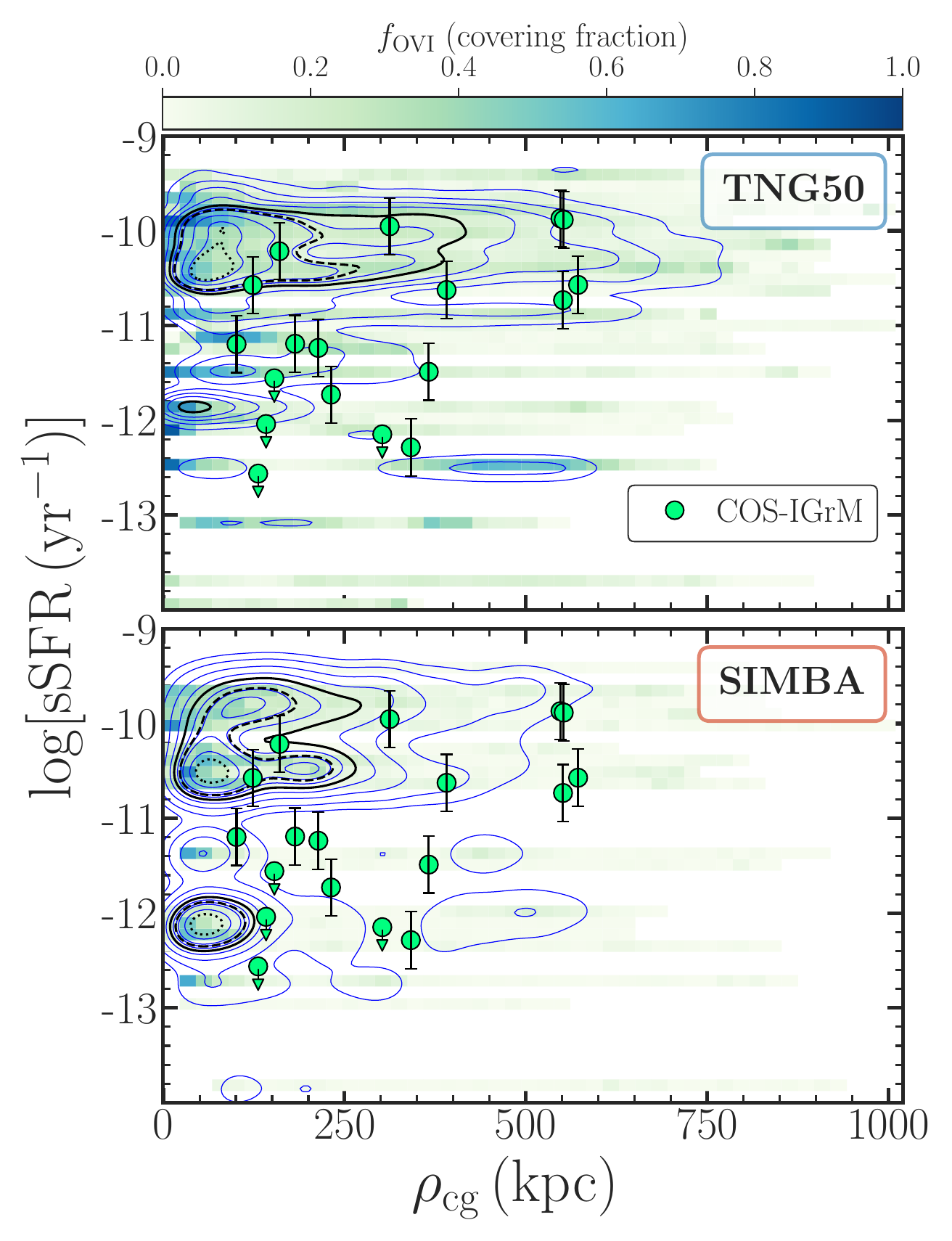}
    \end{minipage}

    \caption{
    (Left) O\,\textsc{vi} column density (\(N_{\mathrm{OVI}}\)) versus impact parameter (\(\rho_{cg}\))
    from the nearest member galaxy. Top row: TNG50 groups. Bottom row: \textsc{SIMBA} groups.
    Left column: all sight-lines. Centre column: systems with \(\log[\mathrm{sSFR}] \leq -11\).
    Right column: systems with \(\log[\mathrm{sSFR}] > -11\).
    The 2-D KDE and scatter styling match Fig.~\ref{fig:Combined_ColumnDensity_vs_rho_TNG50_SIMBA}.
    (Right) Specific star‐formation rate (sSFR) of the nearest group member galaxy for absorbers,
    plotted against impact parameter to that galaxy. The coloured pixels show covering fraction
    (detected absorbers divided by total sight‐lines per bin), while overlaid 2-D KDE contours trace
    the density of detected absorbers, following the styling of
    Fig.~\ref{fig:Combined_ColumnDensity_vs_rho_TNG50_SIMBA}. Top: TNG50 groups.
    Bottom: \textsc{SIMBA} groups.}
    \label{fig:combined_OVI_and_sSFR_panels}
\end{figure*}

Left panel of Figure~\ref{fig:combined_OVI_and_sSFR_panels} presents O\,\textsc{vi} column densities as a function of distance from the nearest galaxy for TNG50 (Top) and SIMBA (Bottom). We also divide our sample into two subsets: (1) sightlines closest to a non-star-forming galaxy (\(\log \text{sSFR} \leq -11\)) as shown in the centre panels  and (2) sightlines closest to a star-forming galaxy (\(\log \text{sSFR} > -11\)) presented in right panels . For TNG sample, we find that among high column density absorbers (\(\log N_{\text{OVI}} > 15\)), \(86\%\) are located within \(\rho < 200 \text{ kpc}\) of a star forming galaxy, suggesting a possible physical connection between these absorbers and their nearest galaxy.  In the star-forming case, \(1.62\%\) of the detected absorbers are high column density detections, whereas in the non-star-forming case, this fraction drops to \(0.79\%\). Overall, among the total \(5676\) absorbers with \(\log N_{\text{OVI}} > 15\), \(90\%\) originate from star-forming galaxies, indicating a possible correlation between O\,\textsc{vi} absorption and ongoing star formation. Among all detections with \(\log N_{\text{OVI}} < 15\), approximately 17\% are associated with galaxies having $\log(\mathrm{sSFR}) \leq -11$, while the remaining 83.0\% are linked to galaxies with $\log(\mathrm{sSFR}) > -11$.

Right panel of Figure~\ref{fig:combined_OVI_and_sSFR_panels} shows the sSFR of the galaxy nearest to O\textsc{vi} absorbers as a function of the impact parameter ($\rho_{\mathrm{cg}}$) from the closest galaxy. The colored pixels represent the fraction of detected absorbers to the total absorbers in the $(\rho_{\rm cg},\log{\rm sSFR})$ bin, while the overlaid contours trace the overall smoothed density of detected absorbers alone. 
For TNG sample, we find that the O \textsc{vi} covering fraction is highest at small impact parameters, exceeding \(f\approx0.5\) for \(\rho_{\rm cg}\lesssim100\)\,kpc.  Beyond \(\rho_{\rm cg}\sim250\)\,kpc, the covering fraction declines smoothly.  Most detections are concentrated around \(\log\mathrm{sSFR}\approx-10.5\) at low impact parameters, whereas at very low sSFR (\(\log\mathrm{sSFR}\lesssim-12.5\)) we find the O\textsc{vi} covering fraction to be much lesser (\(f<0.1\)). There isn't any significant overlap in the COS-IGrM data as most of the detections and non-detections fall in the moderate to low confidence regions.A similar trend was seen for the SIMBA sample albeit at lower strength than TNG.

Overall, a key distinction between TNG50 and SIMBA is that in TNG50 the bulk of the high column density absorbers (\(\log N_{\mathrm{OVI}} > 15\)) are concentrated within \(\rho \lesssim 100~\mathrm{kpc}\) of star-forming galaxies whereas this is not the case in SIMBA.

\section{Discussion}\label{sec:discussion}

\begin{figure*}[t]
    \centering
    \includegraphics[width=\linewidth]{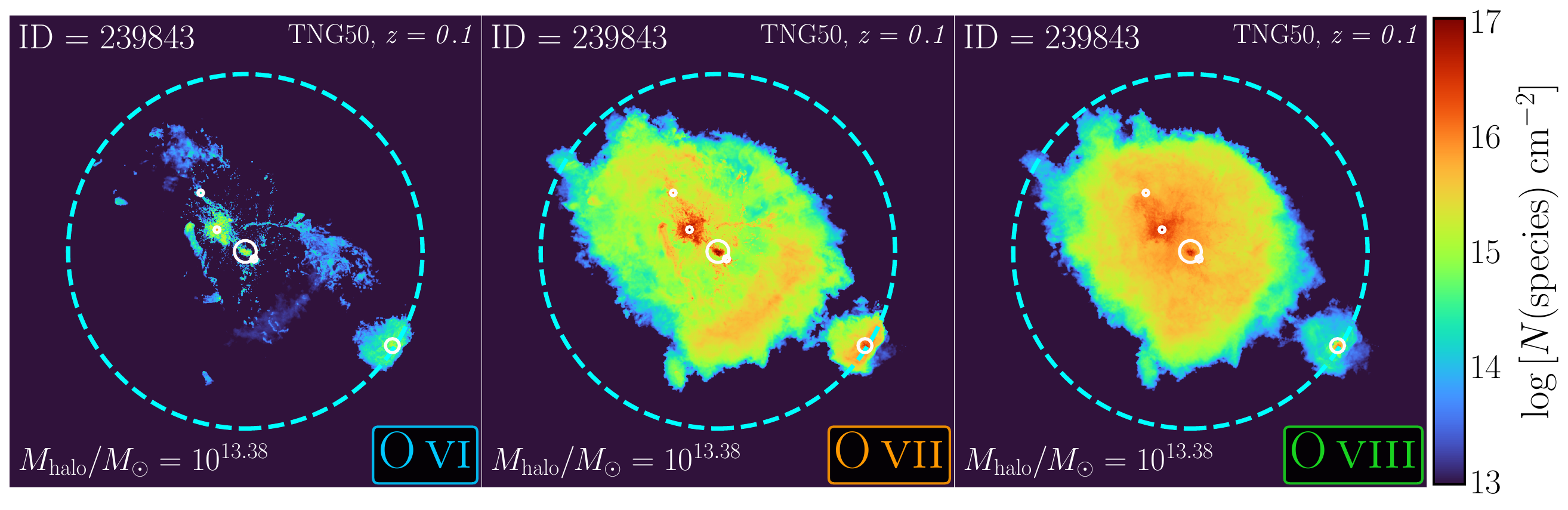}
    \caption{Projected column–density maps of O\,\textsc{vi}, O\,\textsc{vii}, and O\,\textsc{viii} for the \textsc{TNG50} galaxy group \emph{ID 239843}. The dotted cyan circle marks 1.5\,$R_{\mathrm{vir}}$ (660~ckpc/$h$). Smaller white circles indicate member galxies with $L \geq L_{*}$, drawn with radii equal to three times their stellar half–mass radius.}
    \label{fig:single_groups_OVI_OVII_OVII}
\end{figure*}

\begin{figure*}[t]
    \centering

    \begin{minipage}[t]{0.49\textwidth}
        \centering
        \includegraphics[width=\linewidth, trim={0 0 0 0}, clip]
        {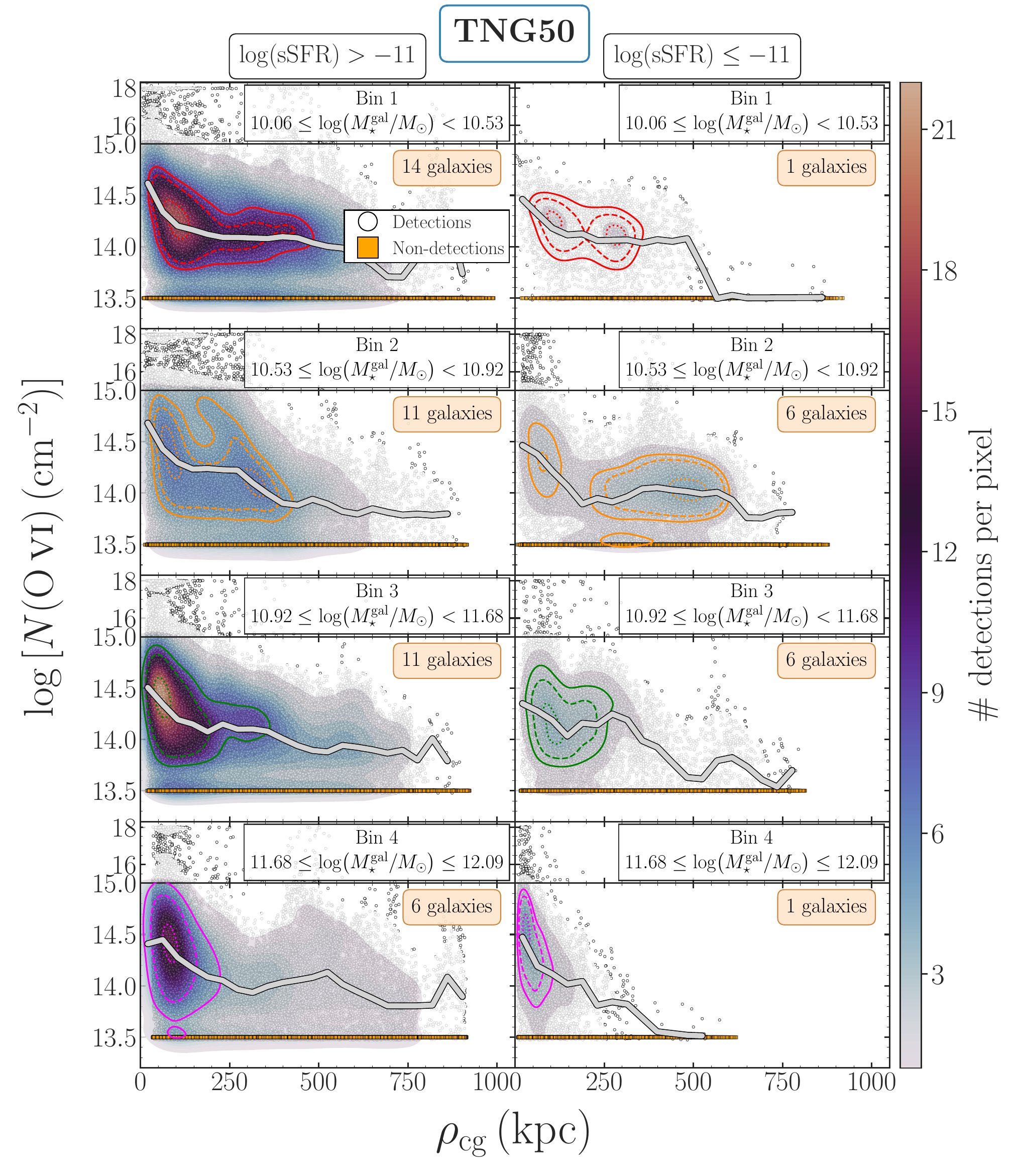}
    \end{minipage}%
    \begin{minipage}[t]{0.49\textwidth}
        \centering
        \includegraphics[width=\linewidth, trim={0 0 0 0}, clip]
        {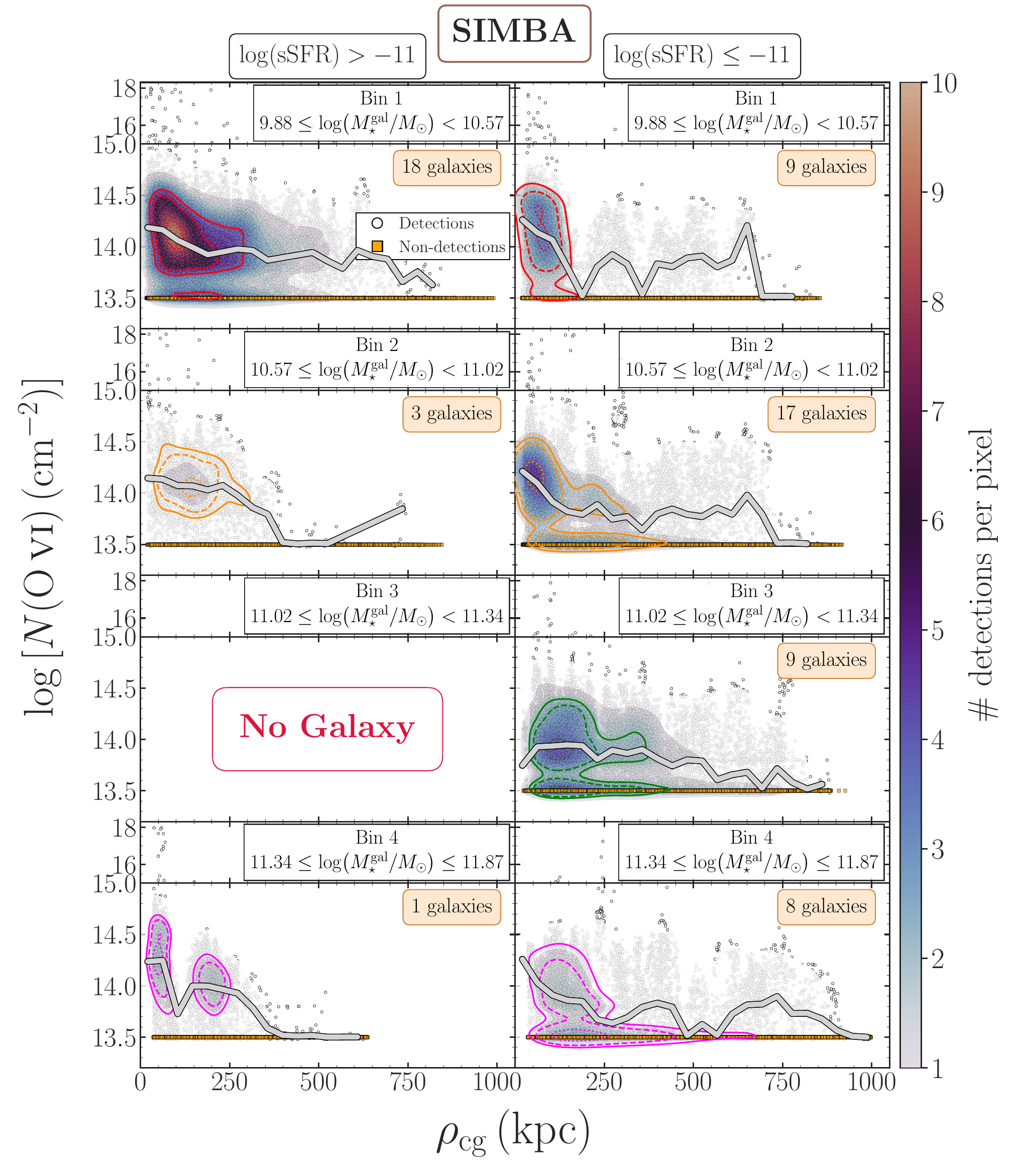}
    \end{minipage}

    \caption{Column density of O\,\textsc{vi} absorbers as a function of impact parameter to the nearest galaxy, $\rho_{\mathrm{cg}}$. Each composite is arranged into four stellar–mass bins (annotated in the top right of each panel) and further divided by star–forming state (columns: left, star–forming; right, non–star–forming). Contours and symbols follow the styling of Fig.~\ref{fig:Combined_ColumnDensity_vs_rho_TNG50_SIMBA}.}
    \label{fig:logN_vs_rho_by_stellar_mass_combined}
\end{figure*}

We find a significant discrepancy between the O\textsc{vi} covering fractions ($f_{\text{OVI}}$) measured from the COS-IGrM survey ($44\pm5\%$) and those predicted by the TNG50 and SIMBA simulations (Figure~\ref{fig:covering_fraction}) when identical synthetic observations are replicated. TNG50 yields an $f_{\text{OVI}}$ of \(20.62 \pm 2.56\%\), whereas SIMBA predicts a significantly lower $f_{\text{OVI}}$  of \(5.98 \pm 0.82 \%\). A similar trend is also observed for the sample of sightlines matched in impact parameter with the COS-IGrM survey. One likely explanation for the low O\,\textsc{vi} covering fraction is that a substantial fraction of oxygen resides in higher ionization states, namely O\,\textsc{vii} and O\,\textsc{viii}. Another plausible cause could be that the overall baryon fraction or metallicity in COS-IGrM sample is higher than the simulated groups.
 
We show projections of three ionization states of oxygen—O\,\textsc{vi}, O\,\textsc{vii}, and O\,\textsc{viii} for a \textsc{TNG50} halo of our sample in Figure~\ref{fig:single_groups_OVI_OVII_OVII}. The maps reveal that most of the oxygen is not in the O\,\textsc{vi} state, but instead resides in the higher ionization states O\,\textsc{vii} and O\,\textsc{viii}. We also see that O\,\textsc{vi} is concentrated in relatively small, clumpy regions, whereas the higher ionization states are distributed in a more diffuse pattern throughout the halo. This trend is consistently observed across all TNG50 groups of our sample, as illustrated in Figures~\ref{fig:projection_tng_OVII} and \ref{fig:projection_tng_OVIII}, in contrast to the more clumpy and localized distribution of O\,\textsc{vi} shown in Figure~\ref{fig:projection_tng}. In the TNG300 and TNG100 simulations, \citet{Nelson_2018_OVI} show a similar trend in the relative abundances of these ions across the full halo population, with oxygen increasingly dominated by higher ionization states in group-mass systems.

The differences between simulations likely originate from their distinct feedback implementations and resolutions. TNG50 employs dual-mode AGN feedback characterized by thermal and kinetic energy injection, facilitating redistribution of metals in the IGrM \citep{10.1093/mnras/stw2944,10.1093/mnras/stx3112} whereas, SIMBA's jet-driven kinetic feedback is efficient at expelling gas beyond the halo, leading to a deficiency of the O\,\textsc{vi}-bearing gas within $1.5R_{vir}$ \citep{Dave_2019}.

Moreover, the numerical resolution of the two simulations differs substantially. In \textsc{SIMBA}, typical gas element masses that are \(\sim 100\) times larger than in \textsc{TNG50}, potentially limiting \textsc{SIMBA}'s ability to resolve the small-scale, multiphase structure that can host O\,\textsc{vi}-bearing gas in group haloes \citep{Dave_2019}. In this work we do not attempt to isolate the impact of this resolution difference from that of the distinct feedback implementations; our comparison therefore reflects the combined effects of both. Resolution studies that systematically vary CGM resolution at fixed subgrid physics have shown that absorber column densities and covering fractions in the CGM can be strongly resolution dependent \citep[e.g.,][]{Hummmels_2019,Peeples_2019,Van_2019,Tortora_2024}, and a similar controlled experiment for group-mass haloes in the \textsc{TNG} and \textsc{SIMBA} models would be required to establish whether numerical resolution plays a dominant role in setting the O\,\textsc{vi} content of the IGrM. 

The impact of resolution can be seen within the various TNG50 runs. The O\,\textsc{vi} covering fraction decreases with decreasing resolution; however, the variation between adjacent resolution runs (e.g., TNG50-1 and TNG50-2) is comparable to the dispersion among groups within a given run. Similarly, the impact of feedback is seen when comparing the \textsc{SIMBA} fiducial run (``s50'') with the AGN-free variant (``s50noagn''), with the AGN-free run producing a higher covering fraction as expected. Moreover, we also note that the adopted UVB model may also affect the predicted O\,\textsc{vi}, particularly in low-density gas where photoionization dominates \cite{Mallik_2023,Mallik_2024,Khaire_2024,Taira_2025}. The bulk of group-scale O\,\textsc{vi} is at least partially collisionally ionised, limiting this sensitivity, but lower-density components may be affected. Additionally, TNG50 and SIMBA employ different hydrodynamics solvers, using \textsc{AREPO} \citep{Springel_2010,Weinberger_2020} and \textsc{GIZMO} \citep{Hopkins_2015} respectively, which can yield different gas morphologies and metal distributions even when identical physics is adopted \cite{Hu_2023,Strawn_2024}. This should also be kept under consideration when interpreting the comparison.

A crucial finding of our study is that high column density O\,\textsc{vi} absorbers ($\log N_{\text{OVI}} > 15$) are predominantly found in the vicinity of star-forming galaxies within the groups in the TNG sample. Specifically, these absorbers preferentially reside near galaxies with specific star formation rates ($\log \text{sSFR} > -11$). The fraction of high column density ($\log N_{\text{OVI}} > 15$) absorbers are twice as many near actively star-forming galaxy compared to non-star-forming ones. In the absorbers with $\log N_{\text{OVI}} > 15$, $90\%$ are associated with star-forming galaxies, while only $10\%$ are found near non-star-forming galaxies. Although these fractions depend on the relative number of sightlines passing near star‑forming versus non–star‑forming galaxies, the overall trend suggests a possible correlation between the observed O\,\textsc{vi} absorption and the ongoing star‑formation activity. However, this correlation could also be an indirect consequence of black hole feedback than of stellar processes alone as in TNG, kinetic-mode AGN feedback drives quenching and simultaneously heats the CGM, reducing the O\,\textsc{vi} content \cite{Nelson_2018_OVI,Weinberger_2018,Zinger_2020}.

On the other hand, the low-column-density O\,\textsc{vi} absorbers ($\log N_{\text{OVI}}<14$) exhibit no clear correlation with sSFR. This aligns with observational studies indicating that low-column-density absorbers can be ubiquitous in group environments, independent of nearby star formation activity \citep{Johnson_2015,Prochaska_2011}. These weak absorbers may possibly represent the O\,\textsc{vi} of ambient IGrM or older outflows or dynamically mixed intragroup gas rather than recent star formation-related enrichment.

The results also reveal that the distribution of O\,\textsc{vi} absorbers does not follow a simple radial gradient within the groups in both the simulations \citep{mulchaey_2000,helsdon_2000}. Figure~\ref{fig:Combined_ColumnDensity_vs_rho_TNG50_SIMBA} illustrates a relatively flat distribution of column densities from $0.6R_{vir}$ to $1.5R_{vir}$ although we do see a small increase of the higher column density at much lower $\rho$ particularly for TNG sample. Such a uniform radial profile suggests that O\,\textsc{vi}-bearing gas in groups is not be simply tracing a radial density gradient within a static, hydrostatic medium.

Figure~\ref{fig:logN_vs_rho_by_stellar_mass_combined} examines how the distribution of O\,\textsc{vi} absorbers depends simultaneously on three factors: the stellar mass of the nearest galaxy, its star–forming state, and the impact parameter to that galaxy. The figure is organized into four stellar–mass bins of the nearest galaxy (increasing from top to bottom)  and further split by star formation (left: star–forming; right: non star–forming). Each stellar–mass bin have similar number of sightlines.

In the TNG50 sample, high column density absorbers 
(\(\log N_{\mathrm{OVI}} \gtrsim 15\)) are strongly concentrated within 
\(\rho_{\mathrm{cg}} \lesssim 100{-}150\) kpc of star–forming galaxies, with this trend becoming more pronounced at higher stellar masses. In contrast, such absorbers are largely absent around non–star–forming galaxies. This indicates that in TNG50 the presence of strong O\,\textsc{vi} is linked to both stellar mass and ongoing star formation of the nearest galaxy. We also note that for weak absorbers, the radial distributions show little variation between star–forming and non–star–forming galaxies across the different stellar–mass bins. We also note a somewhat anomalous case in Bin~2 (non star–forming), where a pronounced concentration of absorbers appears at\(\rho_{\mathrm{cg}} \sim 500~\mathrm{kpc}\). This feature may arise from missing 
galaxy (or galaxies) in our filtering criteria. In SIMBA, the distributions show little to no dependence on either stellar mass or 
star–formation, with fewer high column density absorbers overall and no comparable central concentration. Overall, the decline in O\,\textsc{vi} detection rates at the highest stellar–mass 
bins can largely be attributed to gravitational heating.

For completeness, we also examine whether the stellar mass of groups correlates with the distribution of O\,\textsc{vi} absorbers. In TNG50, we do not find a strong  overall dependence of O\,\textsc{vi} distribution on group stellar mass, although  there is a hint that stronger absorbers (\(\log N_{\mathrm{OVI}} > 15\)) become  more centrally concentrated in higher–mass bins. In SIMBA, the distributions  appear flatter, with absorbers more frequently found at larger impact parameters.  Overall, these correlations are weaker than those observed with the stellar mass and star–forming state of the nearest galaxy. Similarly, we check for a correlation between group halo mass and the distribution of O,\textsc{vi} absorbers as a function of group impact parameter. In both simulations we find no significant variation with halo mass.

Further examining absorber kinematics, we find that in TNG, approximately $98.6\%$ of the detected O\,\textsc{vi} absorbers are gravitationally bound to their host groups, compared to $94.2\%$ in SIMBA, consistent with observational results from COS-IGrM ($83\%$; \citealt{mccabe_2021}). Characteristic velocity peaks in some groups further suggest dynamic associations with galactic outflows or infalling gas, rather than with diffuse, uniformly distributed IGrM material. Similar kinematic signatures have been reported in previous studies, confirming that O\,\textsc{vi} traces gas in both cooling inflows and starburst-driven outflows \citep{bordoloi_2017,Voit_2019}.

\section{Conclusion}\label{sec:conclusion}

In this study, we compared observed O\,\textsc{vi} absorption properties from the COS-IGrM survey with predictions from the IllustrisTNG and SIMBA cosmological hydrodynamic simulations using the synthetic absorption spectroscopy approach. We explored spatial distributions, column densities, kinematics, and correlation to the properties of the group and nearby galaxies of O\,\textsc{vi}-bearing gas within galaxy groups. The key findings of our analysis are as follows:

\begin{enumerate}

    \item We show that for a finite signal-to-noise ratio and spectral resolution there is a strong bias in detectability. The bias is strongest for absorbers at the detection threshold. Therefore the full impact of the bias is dependent on the population distribution w.r.t. the detection limit.
        
    \item Both TNG50 of the IllustrisTNG suite and SIMBA simulations underpredict the observed O\,\textsc{vi} covering fractions ($f_{\text{OVI}}$) in galaxy groups. TNG50 yields a covering fraction of $f_{\text{OVI}} = 20.62 \pm 2.56\%$, while SIMBA predicts $f_{\text{OVI}} = 5.98 \pm 0.82\%$. This is significantly lower than the observed $f_{\text{OVI}} = 44 \pm 5\%$ in COS-IGrM survey.

    \item The kinematic analysis of the absorbers suggests that approximately $98.6\%$ of detected absorbers in TNG50 groups and $94.2\%$ in SIMBA are gravitationally bound. In addition, the presence of absorbers with distinct velocity peaks offset from the systemic velocity suggests possible signatures of dynamical interactions and/or feedback-driven flows.

    \item Both simulations show a largely uniform radial distribution of O\,\textsc{vi} column densities beyond roughly $0.6R_{\mathrm{vir}}$, with a slight enhancement of stronger absorbers ($\log N_{\mathrm{OVI}} > 15$) at smaller impact parameters in TNG50.

    \item The incidence and strength of O\,\textsc{vi} absorbers depend jointly on the stellar mass, star formation, and proximity to a galaxy. In TNG50, about $86\%$ of high column density absorbers ($\log N_{\mathrm{OVI}} > 15$) lie within $200~\mathrm{kpc}$ of the nearest galaxy, and strong systems are preferentially located within $\rho_{\mathrm{cg}} \lesssim 100$–$150~\mathrm{kpc}$ of star–forming galaxies with higher stellar masses. Such strong absorbers are largely absent around non star forming galaxies. In contrast, SIMBA produces fewer strong O\,\textsc{vi} absorbers overall and shows no clear dependence on these parameters.
    
    \item We do not find a strong correlation between the column density distribution of O\,\textsc{vi} absorbers and the stellar mass of their host groups. In TNG50, groups across the full stellar–mass rang show similar radial 
    profiles, with only a modest excess of absorbers at smaller radii 
    (\(\rho \lesssim 0.6R_{\mathrm{vir}}\)), particularly for strong systems. In SIMBA, the overall trend is flatter, with a relative shift toward absorbers being found at 
    larger impact parameters as stellar mass increases. These results suggest that O\,\textsc{vi}–bearing gas is not tracing a smooth radial gradient of the group–scale IGrM, but instead is more likely associated with localized, multiphase structures.

\end{enumerate}

Overall, the O\,\textsc{vi} absorption in the intragroup environments of the simulations used in this study is significantly lower compared to the COS-IGrM survey. This discrepancy suggests that the prescriptions for group-scale processes in the simulations may not be entirely accurate. However, since our analysis considered only a limited halo mass range of approximately 0.7 dex, extending this analysis to higher mass halos and comparing the results with \citet{stocke_2019} (and similarly lower mass halos) could provide additional insights and help test this hypothesis further. Nevertheless, our findings indicate that O\,\textsc{vi} is an important tracer of the volume-filling, virialized bulk component of the intragroup medium, even within simulated group environments of TNG50 and SIMBA.

\begin{acknowledgments}

We thank Miha Cernetic for help in installing \texttt{PYGAD}.

T.S. and S.B. acknowledge support from grant HST-GO-17093 administered by STScI, the NASA ADAP grant No. 80NSSC21K0643, and NSF AAG grants 2108159, 2408050, and 2511242. T.S. also acknowledges support from the Beus Center for Cosmic Foundations at ASU. D.N. acknowledges funding from the Deutsche Forschungsgemeinschaft (DFG) through an Emmy Noether Research Group (grant number NE~2441/1-1).

T.S., S.B., and T.M. acknowledge the land and the native people of the Salt River Valley, on whose ancestral territories Arizona State University’s campuses are located, including the Akimel O’odham (Pima) and Pee Posh (Maricopa) Indian Communities, whose care and keeping of these lands allow us to be here today.

\end{acknowledgments}

\AtBeginEnvironment{thebibliography}{
  \small
  \setlength{\itemsep}{0.2ex}
  \setlength{\parsep}{-4pt}
}

\bibliographystyle{aasjournal}
\bibliography{references}

\FloatBarrier 
\appendix

\setcounter{figure}{0}%
\renewcommand{\thefigure}{A.\arabic{figure}}%

\begin{figure*}
  \centering
  \includegraphics[width=\linewidth]{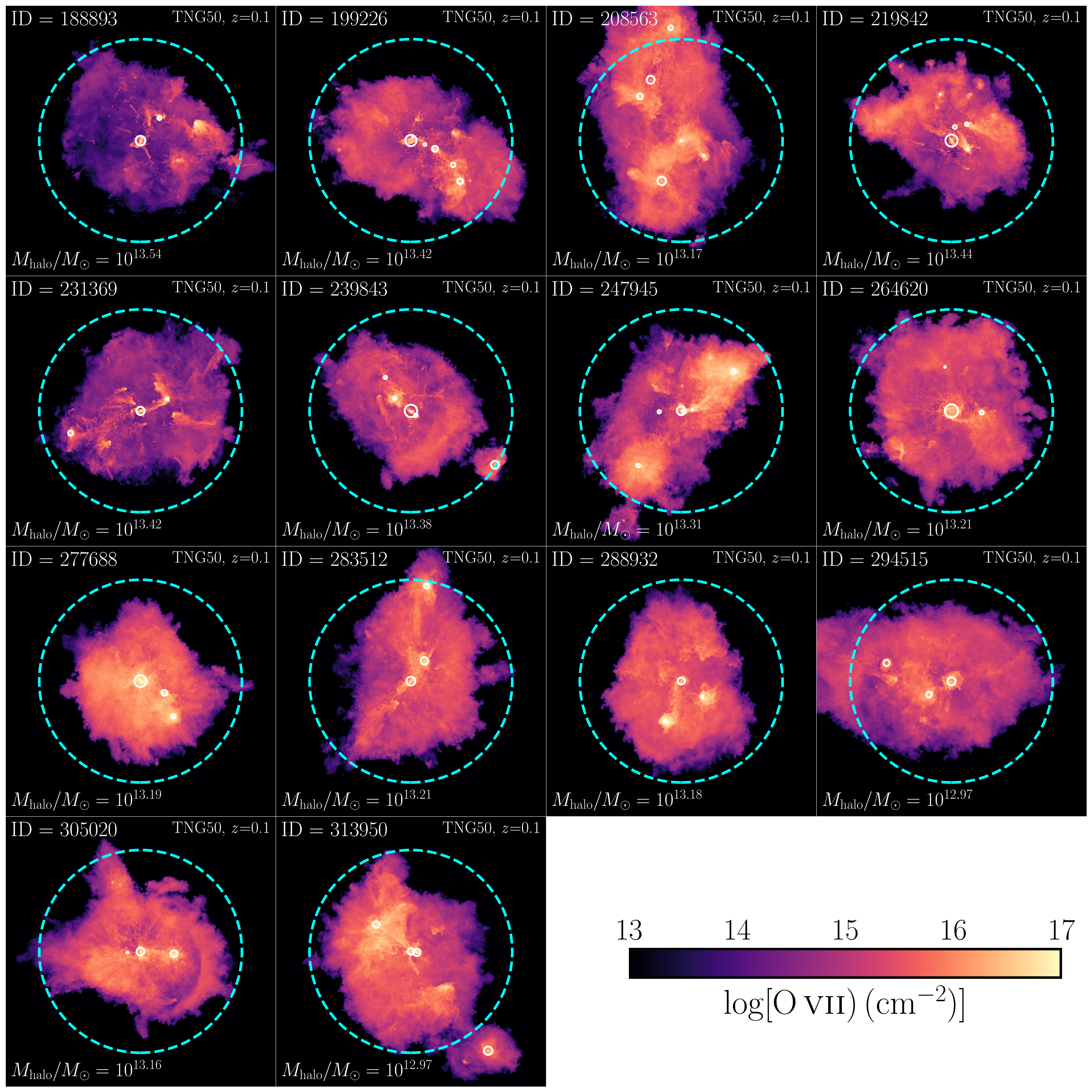}
    \caption{Projected O\,\textsc{vii} column density maps for all selected groups in TNG50 at \( z = 0.1 \), as listed in Table \ref{tng_group_table}. In each panel, the dashed cyan circle has a radius of \( 1.5R_{200c} \), while the box spans \( 4R_{200c} \) along both axes. The ID of the central subhalo of each group is indicated in the top-left corner, and the halo mass is shown in the bottom-left corner. Smaller white circles represent member subhalos with \( L \geq L^* \), where their radii are three times the stellar half-mass radius.}  
    \label{fig:projection_tng_OVII}
\end{figure*}

\begin{figure*}
  \centering
  \includegraphics[width=\linewidth]{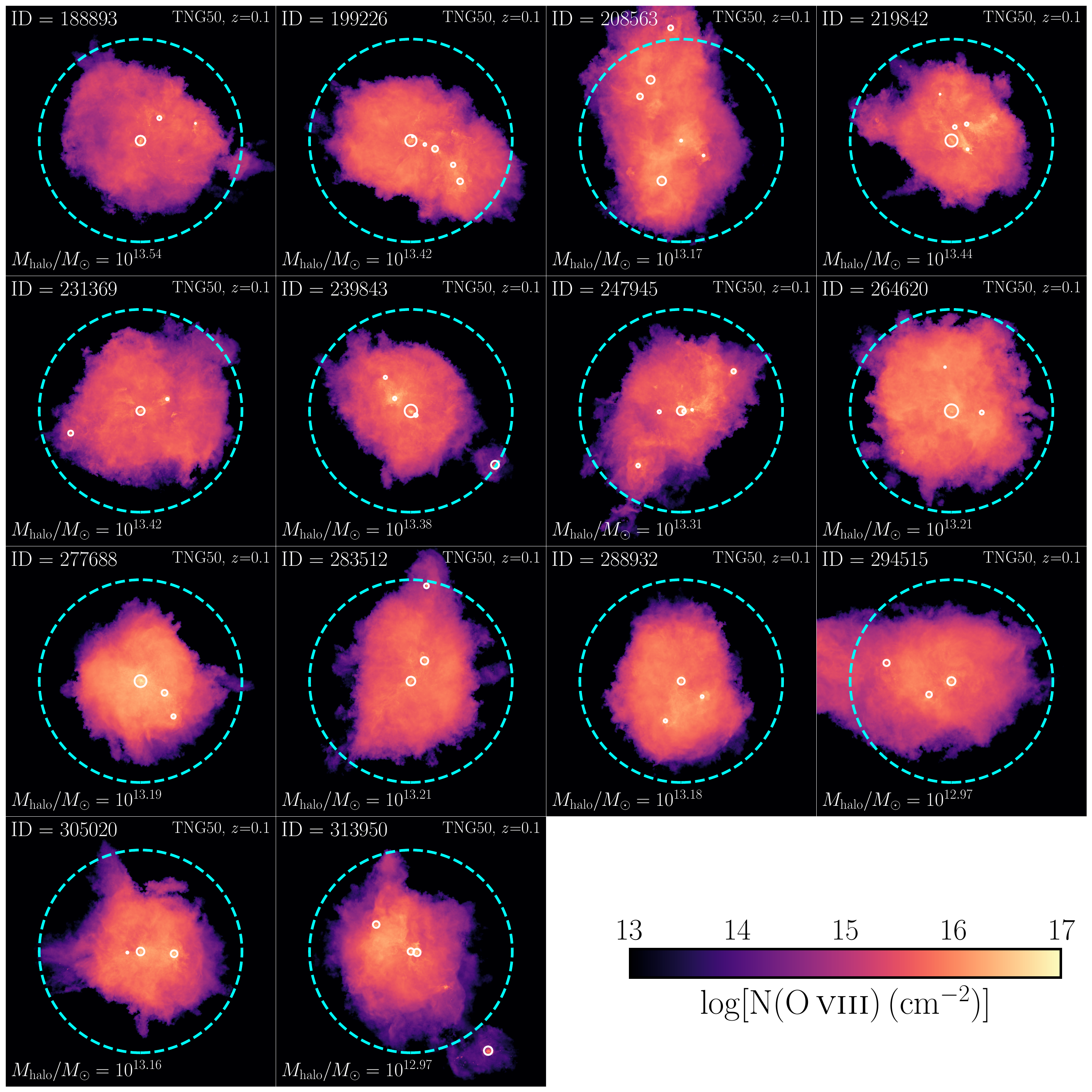}
  \caption{Same as Fig.~\ref{fig:projection_tng_OVII}, but for O\,\textsc{viii}.}
  \label{fig:projection_tng_OVIII}
\end{figure*}

\begin{figure*}
  \centering
  \includegraphics[width=\linewidth]{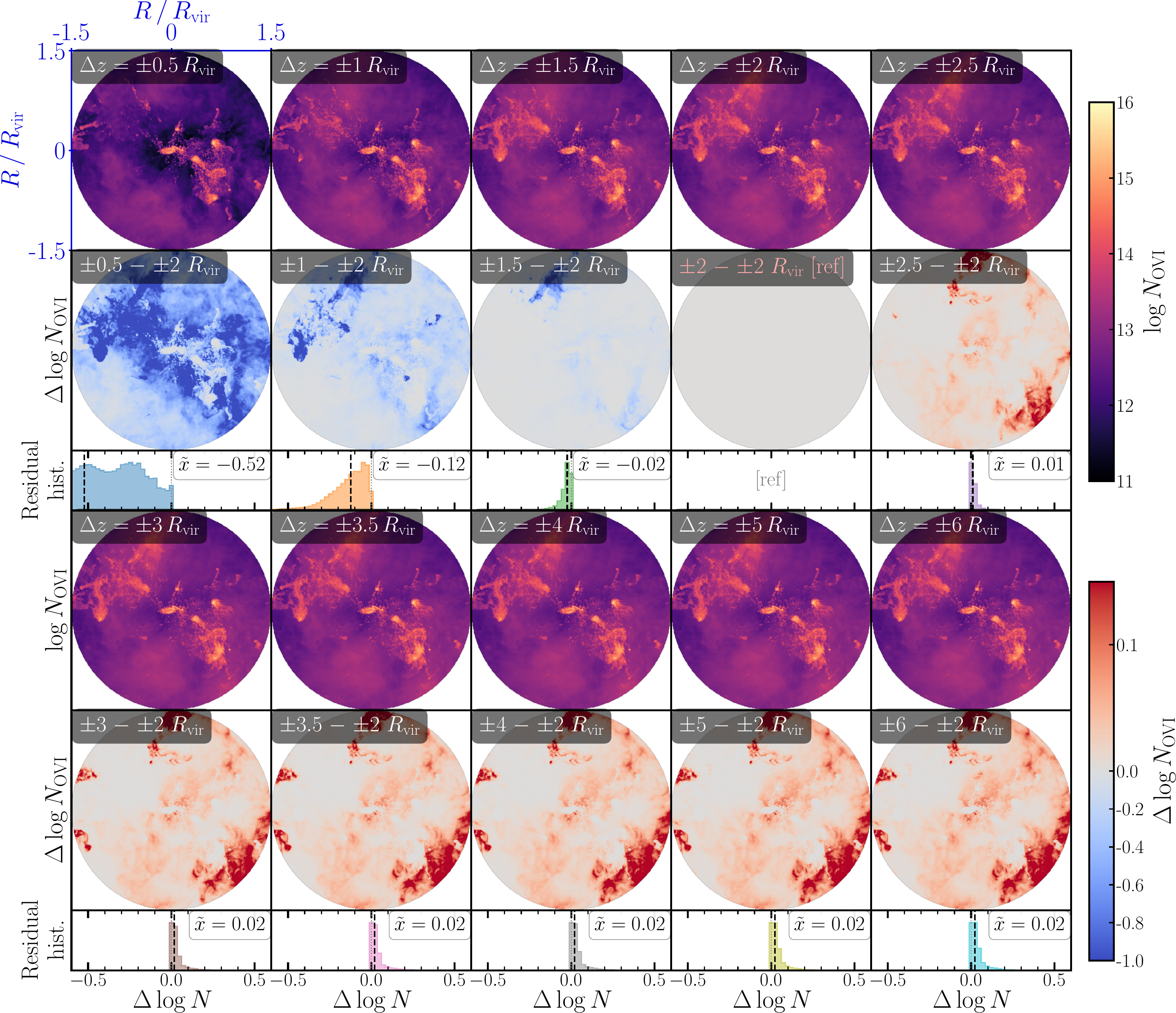}
    \caption{Path-length convergence test for a representative TNG50 halo \textit{(ID=199226)}. Top panels: projected $\log N_{\mathrm{OVI}}$ maps for increasing half-path length $\Delta z_c$. Middle panels: residual maps relative to the $\pm 2\,R_{\mathrm{vir}}$ case. Bottom panels: corresponding residual histograms. The incremental contribution beyond $\pm 2\,R_{\mathrm{vir}}$ is small, indicating convergence.}
  \label{fig:full_panel_path}
\end{figure*}

\end{document}